# A combined ultrasonic flow meter and binary vapour mixture analyzer for the ATLAS silicon tracker.

**R. Bates**[a]**, M. Battistin**[b]**, S. Berry**[b]**, J. Berthoud**[b]**, A. Bitadze**[a]**, P. Bonneau**[b]**,
**J. Botelho-Direito**[b]**, N. Bousson**[c]**, G. Boyd**[d]**, G. Bozza**[b]**, E. Da Riva**[b]**, C. Degeorge**[e]**,
**B. DiGirolamo**[b]**, M. Doubek**[f]**, D. Giugni**[g]**, J. Godlewski**[b]**, G. Hallewell**[c*]**, S. Katunin**[h]**,
**D. Lombard**[b]**, M. Mathieu**[c]**, S. McMahon**[i]**, K. Nagai**[j]**, E. Perez-Rodriguez**[b]**, C. Rossi**[k]**,
**A. Rozanov**[c]**, V. Vacek**[f]**, M. Vitek**[f]** and L. Zwalinski**[b]

[a] *SUPA School of Physics and Astronomy, University of Glasgow,*
  *Glasgow, G62 7QB, United Kingdom*
[b] *CERN,*
  *1211 Geneva 23, Switzerland*
[c] *Centre de Physique des Particules de Marseille,*
  *163 Avenue de Luminy, 13288 Marseille Cedex 09, France*
[d] *The Homer L. Dodge Department of Physics and Astronomy, University of Oklahoma,*
  *Norman, OK 73019, United States of America*
[e] *Indiana University, Department of Physics, 727 East Third St.,*
   *Bloomington, IN 47405-7105, United States of America*
[f] *Czech Technical University in Prague, Department of Applied Physics,*
  *Technická 4, 166 07 Prague 6, Czech Republic*
[g] *INFN Milano and Università di Milano, Dipartimento di Fisica,*
  *Via Celoria 16, 20133 Milano, Italy*
[h] *B.P. Konstantinov Petersburg Nuclear Physics Institute (PNPI),*
  *188300 St. Petersburg, Russia*
[i] *Rutherford Appleton Laboratory - Science & Technology Facilities Council,*
  *Harwell Science and Innovations Campus, Didcot OX11 OQX, United Kingdom*
[j] *Graduate School of Pure and Applied Sciences, University of Tsukuba,*
  *1-1-1 Tennodai, Tsukuba, Ibaraki 305-8577, Japan*
[k] *Department of Mechanical Engineering - Thermal Energy and Air Conditioning Division,*
  *Università degli Studi di Genova Via All'Opera Pia 15a - 16145 Genova, Italy*

   *E-mail:* `gregh@cppm.in2p3.fr`

---

[*] Corresponding author.

ABSTRACT:

An upgrade to the ATLAS silicon tracker cooling control system may require a change from $C_3F_8$ (octafluoro-propane) evaporative coolant to a blend containing 10-25 % of $C_2F_6$ (hexafluoro-ethane). Such a change will reduce the evaporation temperature to assure thermal stability following radiation damage accumulated at full LHC luminosity.

Central to this upgrade is a new ultrasonic instrument in which sound transit times are continuously measured in opposite directions in flowing gas at known temperature and pressure to deduce the $C_3F_8/C_2F_6$ flow rate and mixture composition. The instrument and its Supervisory, Control and Data Acquisition (SCADA) software are described in this paper.

Several geometries for the instrument are in use or under evaluation. An instrument with a 'pinched axial' geometry intended for analysis and measurement of moderate flow rates has demonstrated a mixture resolution of $3.10^{-3}$ for $C_3F_8/C_2F_6$ molar mixtures with ~20 % $C_2F_6$, and a flow resolution of 2 % of full scale for mass flows up to $30 gs^{-1}$. In mixtures of widely-differing molecular weight (mw), higher mixture precision is possible: a sensitivity of $< 5.10^{-5}$ to leaks of $C_3F_8$ into part of the ATLAS tracker nitrogen envelope (mw difference 160) has been seen.

An instrument with an angled sound path geometry has been developed for use at high fluorocarbon mass flow rates of around $1.2 kgs^{-1}$ - corresponding to full flow in a new 60kW thermosiphon recirculator under construction for the ATLAS silicon. Extensive computational fluid dynamics studies were performed to determine the preferred geometry (the ultrasonic transducer spacing and placement together with the sound crossing angle with respect to the vapour flow direction). A prototype instrument with 45º crossing angle has demonstrated a flow resolution of 1.9 % of full scale for linear flow velocities up to $15 ms^{-1}$.

In addition a further variant of the instrument is under development to allow the detection and elimination of incondensable vapour accumulating in the condenser of a fluorocarbon recirculator.

The combined flowmeter and binary gas analysis instrument has many potential applications, including the analysis of hydrocarbons, vapour mixtures for semi-conductor manufacture and anaesthetic gas mixtures.

# Contents

## 1. Introduction

We describe a combined ultrasonic gas mixture analyzer/flow meter for continuous real time composition analysis of binary gas mixtures. The instrument can be used in many applications where knowledge of binary gas composition is required. The combined flow measurement and mixture analysis algorithm employs sound velocity measurements in two directions in combination with measurements of the pressure and temperature of the gas mixture being analyzed.

The instrument – implemented in several geometries described in the following sections - is composed of a pair of ultrasonic transducers mounted within a gas enclosure intended for operation at elevated, atmospheric or sub-atmospheric pressures, together with custom measuring electronics and analyzing software.

The development is motivated by a possible future upgrade of the ATLAS silicon tracker evaporative cooling system, in which the currently-used *octafluoro-propane* fluorocarbon evaporative coolant (R218, $C_3F_8$; molecular weight = 188) [1] will be blended with the more volatile *hexafluoro-ethane* (R118, $C_2F_6$; mw = 138). Additionally, the present underground compressor-driven $C_3F_8$ circulation plant will be replaced by a thermosiphon equipped with a surface-mounted condenser, allowing the ~90 m depth of the ATLAS experimental cavern to be exploited to hydrostatically generate the required fluorocarbon liquid delivery pressure. A combined ultrasonic gas mixture analyzer/flow meter will be installed in the thermosiphon vapour return to the surface while an ultrasonic gas mixture analyzer will monitor for un-condensable gas ingress into the sub-atmospheric vapour (headspace) volume of the surface condenser.

An evaporative coolant based on a $C_3F_8/C_2F_6$ mixture will afford the silicon substrates a better safety margin against leakage current-induced thermal runaway [2] caused by cumulative radiation damage as the LHC luminosity profile increases. The temperature of the silicon substrates of the ATLAS Semi Conductor Tracker (SCT) must be maintained at -7 °C or lower [3] while the ATLAS pixel detector modules must be maintained at 0 °C or lower[4], being typically operated around -10 °C.

At full power dissipation, thermal resistances through the support structures attaching the silicon modules to the coolant channels predicate a typical evaporation temperature of ~ -25 °C in the on-detector cooling channels of the 204 SCT and pixel cooling circuits. With pure $C_3F_8$ this temperature corresponds to an evaporation pressure of 1.7$bar_{abs}$. In the highest power cooling circuits (with a dissipation ~288 Watts - corresponding to ~ 6 W per silicon detector module - we are presently unable to achieve in-tube evaporation temperatures (pressures) below ~ -16.5 °C (2.3 $bar_{abs}$) [2] due to excessive pressure drops (in the range 500-800 mbar, depending on power dissipation in the cooling circuit) in inaccessible regions of the exhaust vapour return system. The addition of the more volatile $C_2F_6$ as part of a binary blend will raise the evaporation pressure for the same evaporation temperature - as indicated in table 1 - allowing the pressure drop in the exhaust to be overcome.

**Table 1.** Evaporation pressure at -25°C evaporation temperature in $C_3F_8$ and several $C_3F_8/C_2F_6$ blends.

| Fluorocarbon Coolant | Median (Minimum) Evaporation pressure for evaporation at -25°C | Sound velocity (ms$^{-1}$) in superheated vapour (20°C, 1 $bar_{abs}$) |
|---|---|---|
| $C_3F_8$ | 1.7 (1.7) $bar_{abs}$ | 115.0 |
| 90 % $C_3F_8$/10 % $C_2F_6$ | 2.3 (1.8) $bar_{abs}$ | 116.8 |
| 80 % $C_3F_8$/20 % $C_2F_6$ | 2.7 (2.1) $bar_{abs}$ | 118.7 |
| 70 % $C_3F_8$/30 % $C_2F_6$ | 3.2 (2.3) $bar_{abs}$ | 120.7 |

In table 1 the median and average evaporation pressures are those at which 100 % and 50 % of the injected refrigerant have been respectively transformed to vapour. In the ATLAS evaporative cooling system the coolant flow is fixed (typically to ~120 % of the flow necessary for simultaneous full power operation of all the silicon detector modules in the circuit) and a fraction of the coolant (dependent on the power dissipated in the silicon modules which constitute the "on-detector" evaporator) is evaporated in electrically-powered heaters in the exhaust system (see [1] for more details). The minimum pressures shown in table 1 correspond to those at the outputs of those heaters.

In operation with $C_3F_8/C_2F_6$ blends it will be necessary to monitor the $C_3F_8/C_2F_6$ mixture ratio and if necessary to adjust it with the addition of more of either component. The mixture ratio is best monitored in the warm region of the vapour return tubing (far from the on-detector evaporative cooling channels) where the vapour mixture is in the superheated (single phase) state, simplifying sound velocity predictions. In table 1 the sound velocities at 20°C, 1 $bar_{abs,}$ calculated using the NIST-REFPROP package [5] are shown as examples.

The sound velocity in a binary gas mixture at known temperature and pressure is a unique function of the molar concentration of the two components of differing molecular weight. Ultrasonic binary gas analysis was first used in particle physics for the analysis of the $N_2/C_5F_{12}$ (*dodecafluoro-pentane*: mw = 288) Cherenkov gas radiator of the SLD Cherenkov Ring Imaging Detector [6], as a simpler alternative to on-line refractive index monitoring. Since then it has been adopted in all the major ring imaging Cherenkov detectors, including DELPHI, COMPASS and LHCb**.**

The molar concentration of the two component vapours is determined from a comparison of sound velocity measurements with velocity-composition look-up table data gathered from prior measurements in calibration mixtures or from theoretical derivations made with an appropriate equation of state. The NIST-REFPROP package [5] is based on the most accurate pure fluid and mixture models currently available for a wide range of fluid combinations. It currently implements three models for the thermodynamic properties of pure fluids: equations of state explicit in Helmholtz energy, the modified Benedict-Webb-Rubin (BWR) equation of state, and an extended corresponding states (ECS) model. Mixture calculations employ a model that applies mixing rules to the Helmholtz energy of the mixture components [5]. NIST-REFPROP is presently the optimum for calculating sound velocity in fluorocarbon mixtures - including those containing $C_2F_6$ and $C_3F_8$. It is not, however configured for calculations in mixtures of fluorocarbons with other gases - for example $N_2/C_3F_8$ - as in this work. Optimum sound velocity predictions are presently made using a PC-SAFT[†] equation of state [7]. Sound velocity predictions made with the NIST-REFPROP and PC-SAFT approaches are compared with measurements in calibration mixtures in sections 3.1 and 3.2.

Figure 1 illustrates the variation of sound velocity with molecular weight for a variety of families of gases, including the monatomic (noble) gases, gases with di-atomic molecules, including oxygen ($O_2$) and nitrogen ($N_2$), tri-atomic gases including carbon dioxide ($CO_2$) and hydrocarbons of the alkane ($C_nH_{(2n+2)}$) and alkene ($C_nH_{(2n)}$) families. Also shown are the members of the saturated fluorocarbon ($C_nF_{(2n+2)}$) family of vapours used in particle physics as Cherenkov radiators and evaporative coolants for silicon tracking detectors.

It can be seen from figure 1 that the *log sound velocity - log molecular weight* zone is bounded by two parallel limits. The upper limit represents the monatomic gases with $\gamma$, [the ratio of specific heat at constant pressure to that at constant volume ($C_p/C_v$)] = 1.66. The noble gases lie along this limit. The lower limit is defined by $\gamma = 1$. Gases with progressively more complex chained molecules tend toward the lower limit. It can be seen from figure 1 that while the *log sound velocity - log molecular weight* trajectory follows an index of -0.5 for the noble gases, the index steepens for more complex molecular shapes. This is a manifestation of their deviation from a simple ideal gas formulation of sound velocity, $c$, defined by;

$$c = \sqrt{\frac{\gamma RT}{m}} \qquad (1)$$

where $R$ is the molar gas constant (8,314 J.mole$^{-1}$K$^{-1}$), $T$ is the absolute temperature (K) and $m$ is the molar weight (expressed in kg) - toward more complex formalisms based on more sophisticated equations of state. The "non-ideality" of a gas group is manifested through the variation (decrease) of $\gamma$ with (increasing) molecular weight, under conditions of equivalent temperature and pressure.

Within the ATLAS experiment the described instrument has been used for flowmetry and mixture analysis of $C_3F_8/C_2F_6$ blends and also as a sensitive detector of leaks of the present $C_3F_8$ evaporative coolant into the ATLAS pixel detector nitrogen envelope. These applications are discussed in sections 3.1 and 3.2.

## 2. Principle of operation

Figure 2 illustrates the principle of operation of the instrument. In custom microcontroller-based electronics (section 4), the transmitting transducer is excited with a short burst of high voltage 50 kHz square wave pulses generated in a driver circuit from a TTL precursor pulse train, itself generated in the microcontroller. The receiving transducer is connected to a DC biasing circuit followed by an amplifier and comparator. A fast (40 MHz) transit time clock, generated in the same microcontroller, is started in synchronism with the rising edge of the first transmitted 50 kHz sound pulse. The first received sound pulse crossing the comparator threshold level stops this clock (figure 2). The time between the transmitted and first received sound pulses is measured by the microcontroller, which also handles communication via a USB/RS232 interface.

---

[†] Perturbed Chain – Statistical Associating Fluid Theory

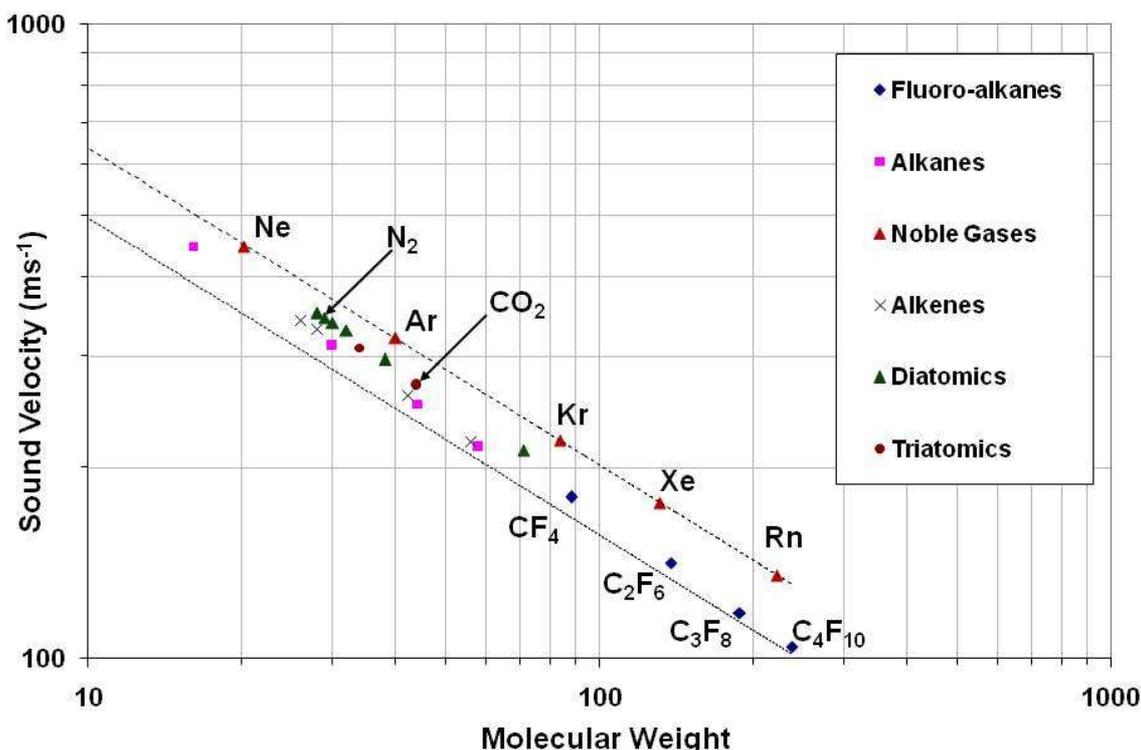

**Figure 1.** Variation of sound velocity with molecular weight for a variety of families of gases, at temperature 295 K and pressure 1 bar$_{abs}$.

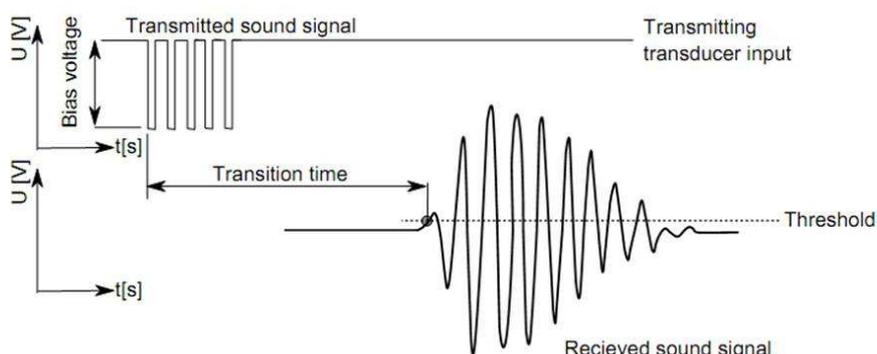

**Figure 2.** Principle of measurement of transit time between the first transmitted sound pulse and the first over-threshold detected pulse.

Figure 3 illustrates the 50 kHz capacitative ultrasonic transducer, originally developed during the 1980s for the Polaroid autofocus instant camera and now marketed by SensComp as the Model 600 series instrument grade ultrasonic transducer. The transducer has a diameter of 42.9 mm and comprises a gold coated mylar foil stretched over a metallic backing plate etched with a spiral groove. The backing plate is biased and excited at a high voltage (80-360 V) with the case and foil grounded. Although the transducer was originally developed for operation in air at ambient temperature and atmospheric pressure, the spiral groove allows gas to access both sides of the diaphragm, allowing operation over a wide pressure range. We have tested and operated these transducers in non corrosive gases at pressures ranging from 5 $bar_{abs}$ to sub-atmospheric pressures of around 100 $mbar_{abs}$ and temperatures in the range -30 to +45 °C.

In the custom electronics sound bursts are sent in opposite directions which may be aligned with - or at an angle - to the gas flow. Rolling average transit times, temperature and pressure data continually stream from a FIFO memory to a supervisory computer presently via RS232, USB, and in future also by CANbus.

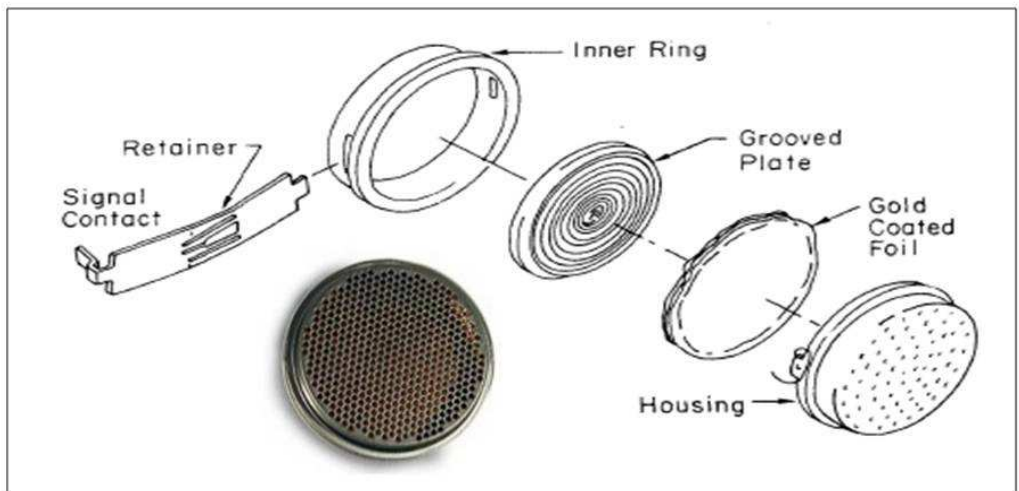

**Figure 3.** The SensComp 50 kHz capacitative ultrasonic transducer, originally developed during the 1980s for the Polaroid autofocus instant camera.

The gas mixture and flow rate are continuously calculated using SCADA software implemented in PVSS-II [8]. Transit times in both directions are used to compute the gas flow rate while the averages of the sound velocities in the two directions are used together with temperature and pressure to compute the binary gas composition by comparison with stored *velocity vs. concentration* look-up tables (Section 6). These tables may be created from prior measurements in calibration mixtures or from theoretical thermodynamic calculations. In future versions of the instrument these calculations may be made in an on-board microcontroller.

## 3. Mechanics, fluid dynamics studies and measurements

Three mechanical implementations of the flow meter/analyzer have been constructed and evaluated:
(i) a gas mixture analyzer/in line flow meter for gas flows up to 250 $lmin^{-1}$, using a "pinched axial" geometry;
(ii) a gas mixture analyzer sampling gas streams at low flow rates around 100 $cm^3 min^{-1}$;
(iii) a gas mixture analyzer/in line flow meter for vapour mass flows around 1.2 $kgs^{-1}$ (335 $ls^{-1}$ in 70 %$C_3F_8$/30 %$C_2F_6$: density 3.58 $kgm^{-3}$ or 307 $ls^{-1}$ in pure $C_3F_8$: density 3.90 $kgm^{-3}$) at 20 ºC and 500 $mbar_{abs}$: characteristic of the temperature and pressure conditions in the warm zone of the vapour return tubing in the future 60 kW ATLAS thermosiphon installation (section 6). Such high flow rates require the use of an angled crossing geometry.

These implementations are described in the following three sections.

## 3.1 "Pinched axial" geometry for analysis & in-line flowmetry for moderate gas flows

The mechanical envelope and ultrasonic transducer mounting are illustrated in figure 4. The transducers are mounted around 660 mm apart in a flanged stainless steel tube of overall length 835 mm. The temperature in the tube is monitored by six NTC thermistors – (100 kΩ at 25 ºC) - giving an average temperature measurement uncertainty of better than ±0.3 ºC. Pressure is monitored with a transducer having ±1 mbar precision.

The transducers are centered and mounted flush with the inboard ends of wide bore sections of tube, between which are welded a pair of diameter reducing cones and a 'pinched' region (of inner diameter 44.3 mm, compared with the transducer diameter of 42.9 mm) through which all the vapour is channelled. Vapour is diverted around the ultrasonic transducers through the use of 5 cm long axial flow-deflecting cones machined from PEEK®.

Three such instruments have been constructed for studies of $C_2F_6/C_3F_8$ blends. Figure 5 illustrates one of these instruments installed in the ATLAS $C_2F_6/C_3F_8$ blend development recirculator. In this simple geometry the vapour flow rate is calculated from the sound transit times measured parallel, $t_{down}$, and anti-parallel, $t_{up}$, to the flow direction, according to the following algorithm:

$$t_{down} = \frac{L}{(c+v)}, \; t_{up} = \frac{L}{(c-v)} \qquad (2)$$

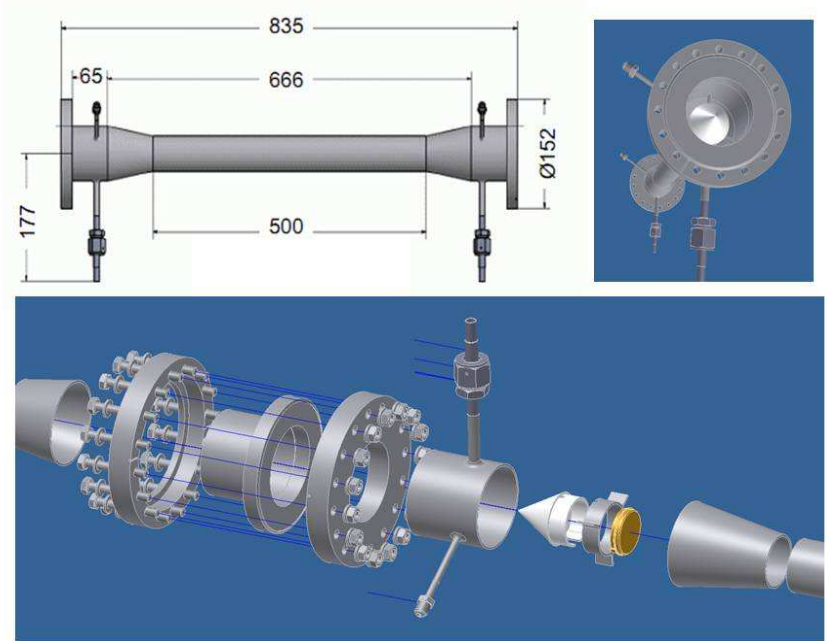

**Figure 4.** The 'pinched axial' mechanical envelope, showing an ultrasonic transducer, its mounting and axial flow-deflecting cone, together with tubes for pressure sensing and the evacuation and the injection of calibration gas.

where $v$ is the linear flow velocity, (ms$^{-1}$), $c$ is the sound velocity in the gas (ms$^{-1}$) and $L$ is the distance (m) between transducers.

The gas volume flow $V$ (m$^3$s$^{-1}$) can then be inferred from the two transit times by:

$$V = \frac{L.A(t_{up}-t_{down})}{2(t_{up}.t_{down})} \qquad (3)$$

where *A* is the internal cross sectional area of the axial flow tube between the ultrasonic transducers (m$^2$).

The sound velocity can also be inferred from the two transit times via:

$$c = \frac{L(t_{up}+t_{down})}{2(t_{up}.t_{down})} \qquad (4)$$

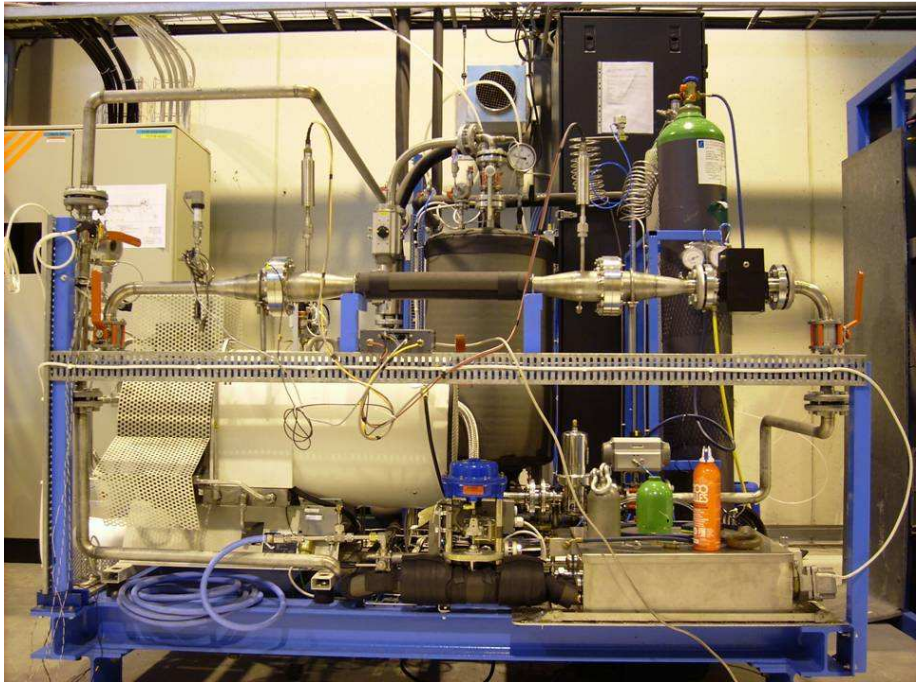

**Figure 5.** 'Pinched axial' geometry analyzer/flow meter installed in the ATLAS C$_2$F$_6$/C$_3$F$_8$ blend, recirculator followed by a Schlumberger Delta G16 integrating gas meter (right foreground). The oil-less piston compressor of the recirculator is shown in the left foreground.

It can be seen from eqs. (2) – (4) that knowledge of the temperature of the gas is not necessary for ultrasonic flowmetry alone.

Figure 6 shows the linearity of the ultrasonic flow meter element of the instrument in C$_3$F$_8$ vapour at 20 °C and 1 bar$_{abs}$ (C$_3$F$_8$ density ~7.9 kgm$^{-3}$) through comparison with a Schlumberger Delta G16 volumetric gas meter (maximum flow rate 25 m$^3$hr$^{-1}$ (417 lmin$^{-1}$), precision ± 1 % of full scale), at flows up to 230 lmin$^{-1}$ (~30 gms$^{-1}$); the maximum mass flow in the presently-available C$_3$F$_8$/C$_2$F$_6$ blend recirculator. At flows above 150 lmin$^{-1}$ the internal rotating piston mechanics of the gas meter became very noisy, having been drained of the normal lubricant charge to avoid contaminating the oil-less compressor of the recirculator.

The horizontal error bars in figure 6 represent the quoted 1 % of full scale error of the gas meter. The vertical error bars reflect the combination of the uncertainty in the main tube diameter (± 0.5 mm), timing precision (± 100ns) and transducer spacing (± 0.1 mm following length calibration in a quasi-ideal static gas, according to the procedure discussed in section 5). The *rms* deviation of the ultrasonic flowmeter relative to the fit (shown as red bands in figure 6) is ± 4.9 lmin$^{-1}$, around 2 % of the full scale flow limit of 230 lmin$^{-1}$.

While the evident linearity allows simple calibration, the deficit from unity slope has been shown from Computational Fluid Dynamics (CFD) to be due to vortex effects in the zones close to the ultrasonic transducers.

The CFD simulation was made using the *OpenFOAM®* package [9] with the SST $k$-$\omega$ [10] turbulence model and a mesh of 545000 polyhedral cells, in order to evaluate the likely effect of the ultrasonic transducers, flow deflection and diameter reduction cones on the measurement precision. "Start condition" volume flows of 100 and 200 lmin$^{-1}$ of $C_3F_8$ at 1 bar$_{abs}$ and 20 °C (density ~ 7.9 kgm$^{-3}$), corresponded to transversally-averaged axial flow velocities of 1.08 and 2.16 ms$^{-1}$ in the 44.3mm ID "pinch" tube linking the transducers. The average velocity in the cylindrical volume defined by the transducer diameter and spacing has been estimated by CFD simulation and assumed to be representative of the velocity seen by the ultrasonic flowmeter. The CFD calculations, shown in figures 7 and 8 and in table 2, suggest a 9% underestimation of the real axial flow velocity due to:

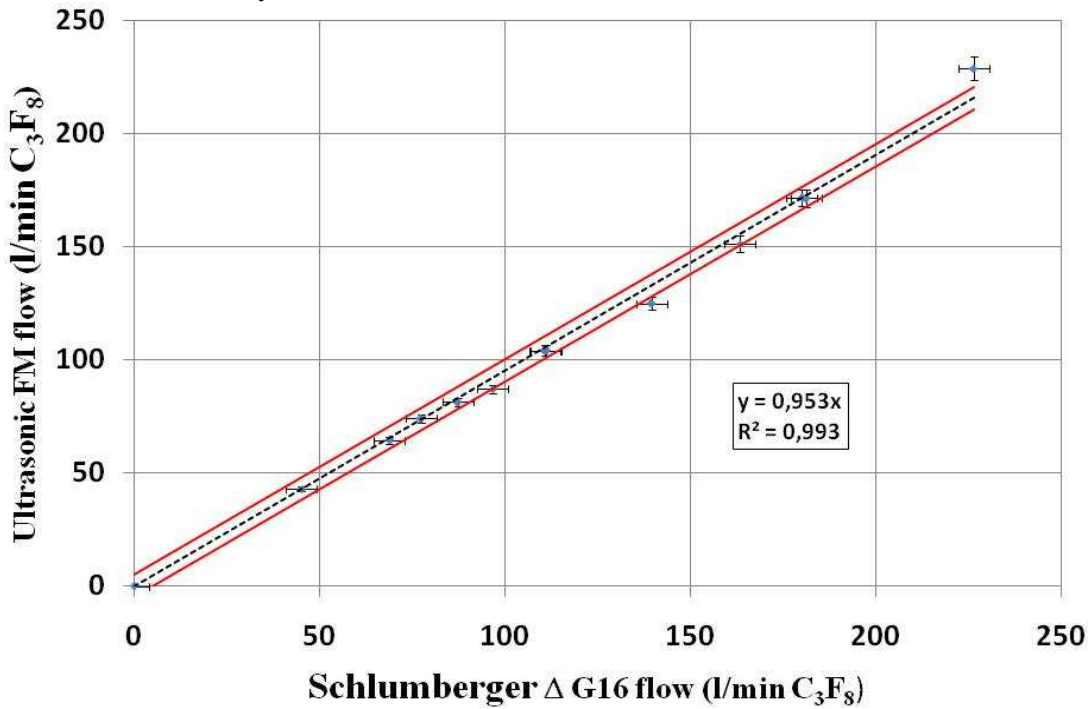

**Figure 6.** 'Pinched axial' ultrasonic flow meter linearity comparison with a Schlumberger Delta G16 gas meter: $C_3F_8$ vapour at 1 bar$_{abs}$ and 20 °C (density ~7.9 kgm$^{-3}$).
The *rms* deviation of the measured points from the fit represents ± 2% of the full flow of 230 lmin$^{-1}$.

- a vortex trailing the upstream ultrasonic transducer within the 83 mm upstream reduction cone (figure 3) where the tube inner diameter shrinks from 72.1 mm to 44.3 mm;
- the slab-like obstruction of the downstream transducer, in front of which the vapour is axially decelerated, then swept transversely to leave the acoustic sensing volume.

**Table 2.** "Pinched Axial" flowmeter: comparison of the expected axial flow velocity in the 44.3 mm "pinch" tube with the CFD-calculated average axial velocity in the cylindrical inter- transducer volume.

| Starting flow rate (l/min) | Expected axial flow velocity in 44.3 mm ID 'pinch' tube (ms$^{-1}$) | CFD volume-averaged velocity between the transducers (ms$^{-1}$) | Deviation from expected average velocity (%) |
|---|---|---|---|
| 100 | 1.08 | 0.983 | -9.14 |
| 200 | 2.16 | 1.95 | -9.65 |

The CFD results of figure 8 and table 2 show the deficit to be constant with flow rate, suggesting that linear calibration against other instrumentation should be possible, as independently demonstrated in figure 6. The deficit suggested from the CFD study is greater than the ~ -4.7 % deficit seen relative to the Schlumberger D16 volumetric gas meter in figure 6. However the gas meter had not been calibrated in the un-lubricated state. The CFD simulations suggest that it is possible also to reduce the deficit by increasing the length of the 44.3 mm diameter tube between the two transducers.

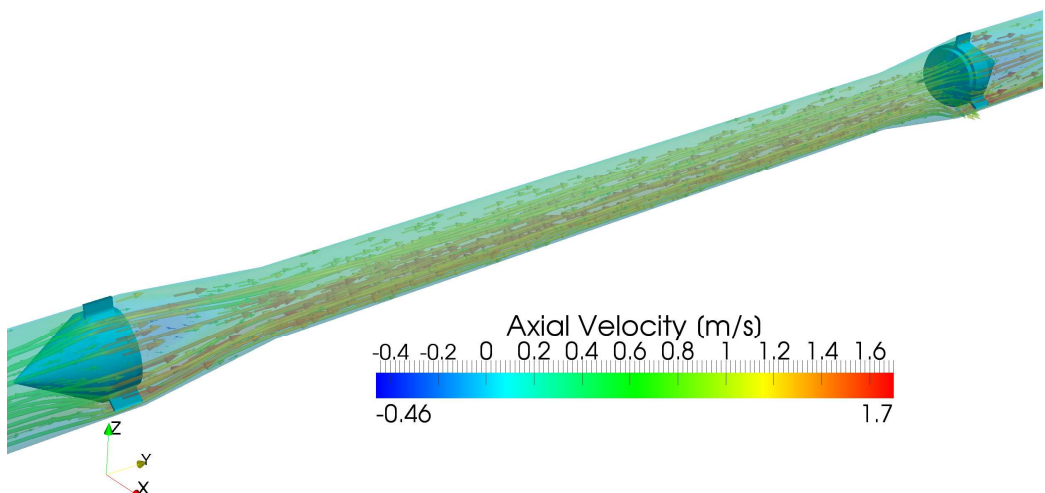

**Figure 7**. Streamlines and glyph vectors in a 3D picture of the pinched flow meter.

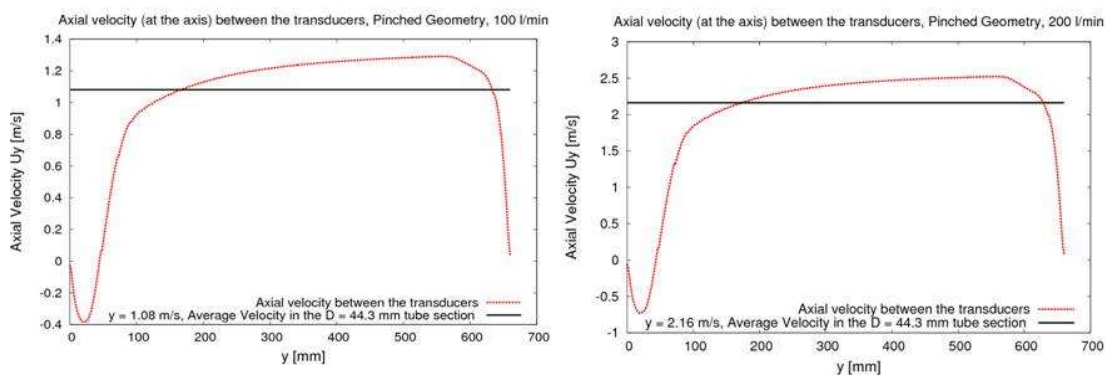

**Figure 8.** Axial flow velocity ($C_3F_8$ at 1 bar$_{abs}$, 20 °C) along the 660 mm axis between the ultrasonic transducers. The average velocities in the 44.3 mm diameter tube of 1.08 ms$^{-1}$ (left plot) and 2.16 ms$^{-1}$ (right plot) respectively correspond to volume flow rates of 100 and 200 lmin$^{-1}$.

Figure 9 illustrates the sound velocity measured in the instrument in varying $C_2F_6/C_3F_8$ molar mixing ratios at a temperature of 19.2 ºC and a pressure of 1.14 bar$_{abs}$.

The mixtures of $C_3F_8$ and $C_2F_6$ shown in figure 9 span the region of thermodynamic interest to the ATLAS silicon tracker cooling application and were set up by partial pressure ratio in the previously-evacuated tube, to create molar mixtures. Component partial pressures were set using a MKS "Baratron" capacitative absolute pressure gauge with a precision of ± 1 mbar. The transducer foil inter-distance had been previously established using the gas calibration procedure described in section 5, to a precision of ± 0.1 mm.

Individual contributions (shown in parentheses below) to the overall 0.05 ms$^{-1}$ sound velocity measurement error (equivalent to 0.042 % at 20 % $C_2F_6$ concentration) in our calibrated mixtures were due to:

- ± 0.2 ºC temperature stability in the sonar tube (equivalent to ± 0.044 ms$^{-1}$);
- ± 4 mbar pressure stability in the sonar tube (± 0.012 ms$^{-1}$) with blend circulation machine in operation;

- ± 0.1 mm transducer inter-foil measurement uncertainty (± 0.018 ms$^{-1}$);
- ± 100 ns electronic transit time measurement uncertainty (± 0.002 ms$^{-1}$).

The average difference between measured and the NIST-REFPROP [5] and PC-SAFT [7] predicted sound velocities in mixtures with (0➔35 %) $C_2F_6$ in $C_3F_8$ were 0.05 % and 0.25 % respectively at pressures around 1 bar$_{abs}$ and temperatures around 19 °C. It is recognized that the present version of NIST-REFPROP [5] is more mature than the PC-SAFT equation of state and is thus the more precise in predicting the thermophysical properties of mixtures of saturated fluorocarbons (having molecular structures of the form $C_nF_{(2n+2)}$).

The precision of mixture determination, $\delta(mix)$, at any concentration of the two components is given by;

$$\partial(mix) = \frac{\partial c}{m} \qquad (5)$$

where *m* is the local slope of the sound velocity/ concentration curve and $\delta c$ is the uncertainty in the sound velocity measurement. For example, in a blend of 20 % $C_2F_6$ in $C_3F_8$ - the sound velocity uncertainty of 0.05 ms$^{-1}$ yields a concentration uncertainty ~0.3 % at 20 % $C_2F_6$, where the slope of the velocity/ concentration curve is ~0.18 ms$^{-1}$%$^{-1}$.

The precision of the sound velocity determinations of the NIST-REFPROP package [5] for $C_2F_6$ and $C_3F_8$ has been independently estimated in [11], as part of an extensive set of comparisons between calculated thermodynamic and thermophysical parameters and measurement data from a large number of source references for 20 refrigerant fluids. The NIST-REFPROP determinations of sound velocity in $C_2F_6$ ($C_3F_8$) superheated vapour over the temperature & pressure ranges 210-475 K & 0.8-32 bar$_{abs}$ (245-342 K & 0.1-16.2 bar$_{abs}$), based respectively on the thermodynamic measurement data of 17 and 13 groups, were compared with sound velocity measurements made for each vapour by 2 groups. In the case of $C_3F_8$ the average deviation was around 1 %, while for $C_2F_6$ the average deviation was around 0.2 %.

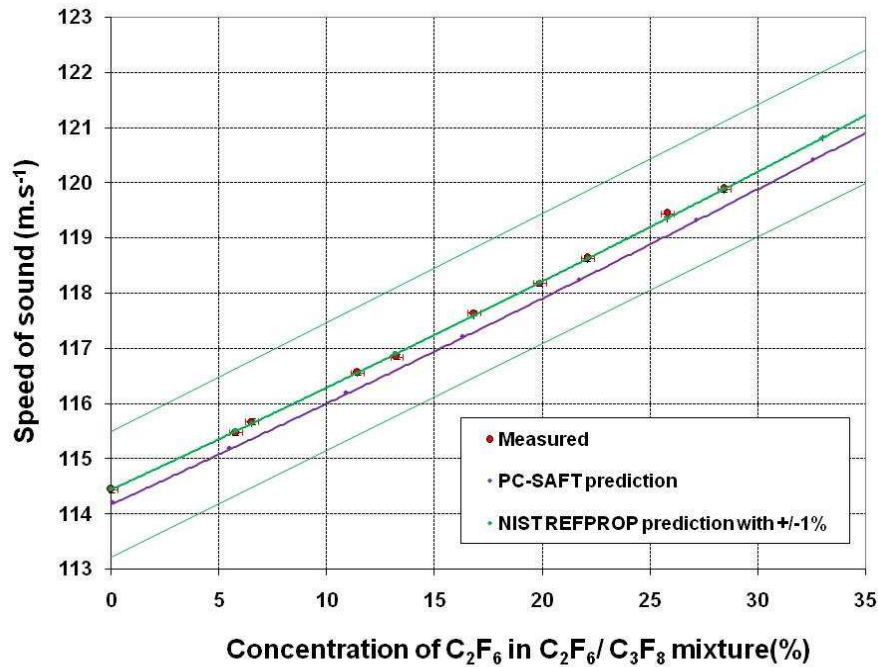

**Figure 9.** Comparison between measured sound velocity data and theoretical predictions in molar $C_3F_8/C_2F_6$ mixtures of thermodynamic interest, at 1.14 bar$_{abs}$ and 19.2°C. NIST-REFPROP sound velocity predictions shown within ± 1 % band, after [11]. The binary gas mixture uncertainty of 0.3% is illustrated in red.

No comparisons have made between sound velocity measurements and predictions from NIST-REFPROP or PC-SAFT for $C_2F_6/C_3F_8$ molar mixtures until this work. Since $C_3F_8$ is the dominant component in the mixture figure 9 illustrates the 1 % sound velocity band (in green) from [11]. Clearly NIST-REFPROP sound velocity predictions are very much closer to our measurements than this. Indeed, figure 9 illustrates the present maturity of NIST-REPROP for the category of saturated fluorocarbons. In the case of our measurements we show the 0.3 % mixture uncertainty (red error bars) coming from our sound velocity measurement uncertainty of 0.05 $ms^{-1}$. We do not show error bars for PC-SAFT predicted sound velocities although it seems that the 0.25 % systematic difference with our measurements is likely to reflect that uncertainty given the present level of maturity of this equation of state.

**3.2 Gas analyzer for sampling gas streams at very low flow rates**

In a second application related to the ATLAS evaporative cooling system, we use ultrasonic binary gas analysis to detect low levels of $C_3F_8$ vapour leaking into the $N_2$ environmental gas surrounding the ATLAS pixel tracker. Gas is aspirated from the pixel tracker volume through an ultrasonic analysis tube with a vacuum pump, as shown in figure 10. A low flow of around 100 ml.min$^{-1}$ is achievable through the (150 m long, 8 mm inner diameter) aspiration tube, with the analyzer tube operating at a pressure of 985 mbar$_{abs}$ ± 2 mbar. Exhaust gas from the analysis tube (typically containing > 99.9 % $N_2$) is vented to an air extraction system for return to the surface. The analyzer has been in continuous operation since January 2010, and is located - together with its data acquisition and supervisory computer - in a subterranean technical cavern alongside the ATLAS experimental cavern, whose floor is 92 m below the surface.

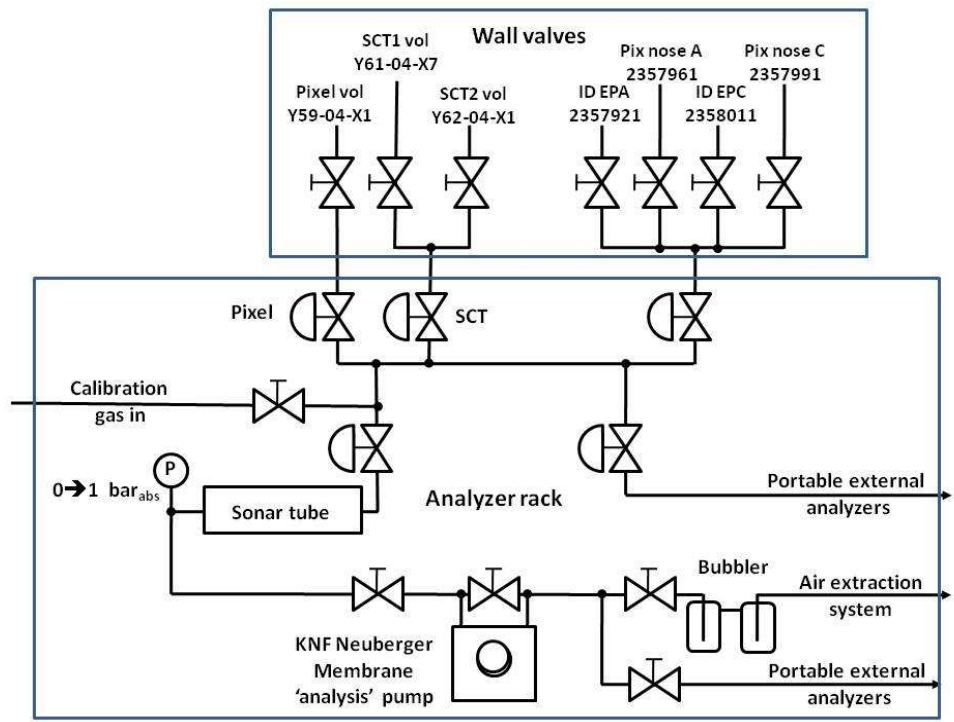

**Figure 10.** Ultrasonic gas mixture analyzer for continuous monitoring of $C_3F_8$ coolant leak into the $N_2$ volume surrounding the pixel detector of the ATLAS silicon tracker.

A valve selection panel allows future expansion to allow the envelope gases of the ATLAS silicon microstrip tracker (SCT) and inner detector volumes to be monitored. Additional analysis tubes are foreseen in the same rack to allow this, together with full automation of gas stream selection.

Figure 11 compares measured sound velocities in molar mixtures of up to 10 % $C_3F_8$ in $N_2$ with sound velocity predictions from the PC-SAFT equation of state [7]. The data of figure 11 are abstracted from measurements and predictions made over the full concentration range from pure $C_3F_8$ to pure $N_2$. The PC-SAFT equation of state was used as NIST-REFPROP is not presently configured for calculations of mixtures of saturated fluorocarbons with $N_2$.

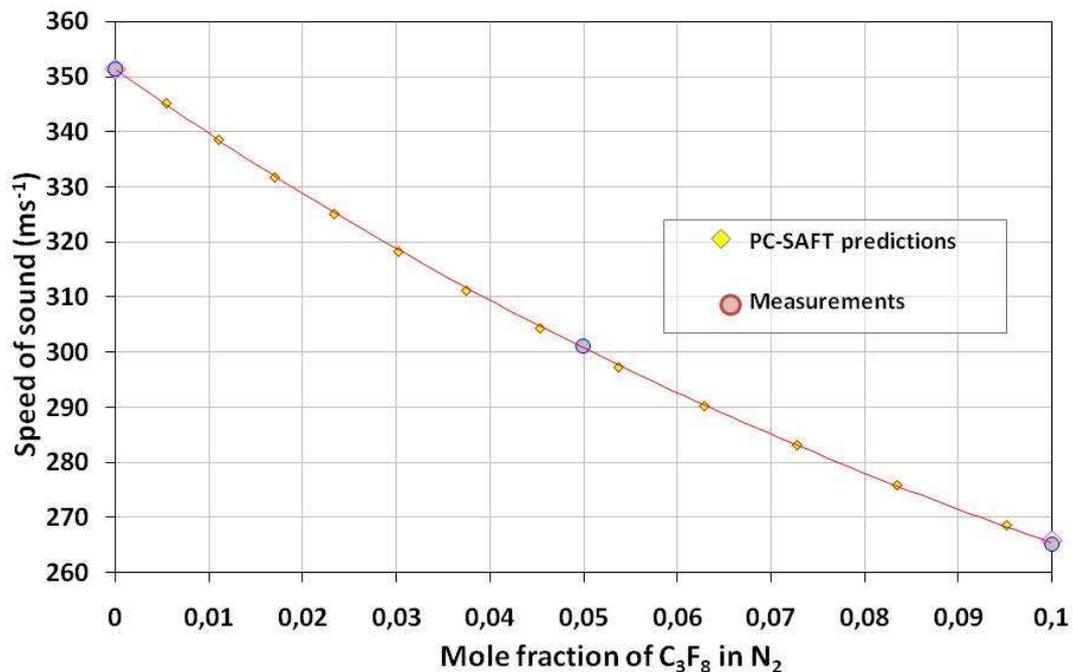

**Figure 11.** Comparison of sound velocity measurements (°) and PC-SAFT predictions (◊) in $C_3F_8/N_2$ mixtures at 1bar abs and 25 ºC. Mixture resolution is ~ $4.10^{-4}$ for concentrations of <1% $C_3F_8$ in $N_2$.

A typical reduction in sound velocity of 0.86 ms$^{-1}$ from that of pure nitrogen is typically observed when the full pixel detector cooling system of 88 independent circuits is fully operating. From the ~-12.27 ms$^{-1}$%$^{-1}$ average gradient of the sound velocity-concentration curve for $C_3F_8$ concentrations in the range 0→0.5 % (figure 11) this sound velocity difference indicates, via eq. (5), a $C_3F_8$ leak ingress of 0.07 % (figure 12). It is evident from comparison of figures 9 and 11 that the gradient of the sound velocity-composition curve (and consequently the resolution in mixture determination) improves considerably with increasing difference in molecular weights of the two components. Thus for $N_2/C_3F_8$ the intrinsic sound velocity uncertainty of around ±0.05 ms$^{-1}$ (section 3.1) correspondingly yields a mixture uncertainty of ±0.004 %.

Figure 12 illustrates a 1-year continuous log of the $C_3F_8$ contamination of the pixel detector environmental $N_2$ envelope. Fluctuations of the measured $C_3F_8$ contamination are correlated with the development of leaks in some of the 88 individual cooling circuits, which have been identified by progressive turn-on or turn-off, but depend also on the purge rate of the $N_2$ gas around the pixel detector, which is typically around 1 m$^3$hr$^{-1}$. Figure 13 illustrates the $C_3F_8$ contamination increase seen in the progressive turn on of cooling circuit groups following the 2011-2012 new-year shutdown of the detector. Though none of the leaks is presently considered severe enough to require the related cooling circuit (and the corresponding section of the ATLAS pixel detector) to be shut down, figure 13 illustrates how the sonar analyzer can be used to rapidly identify a leaking circuit.

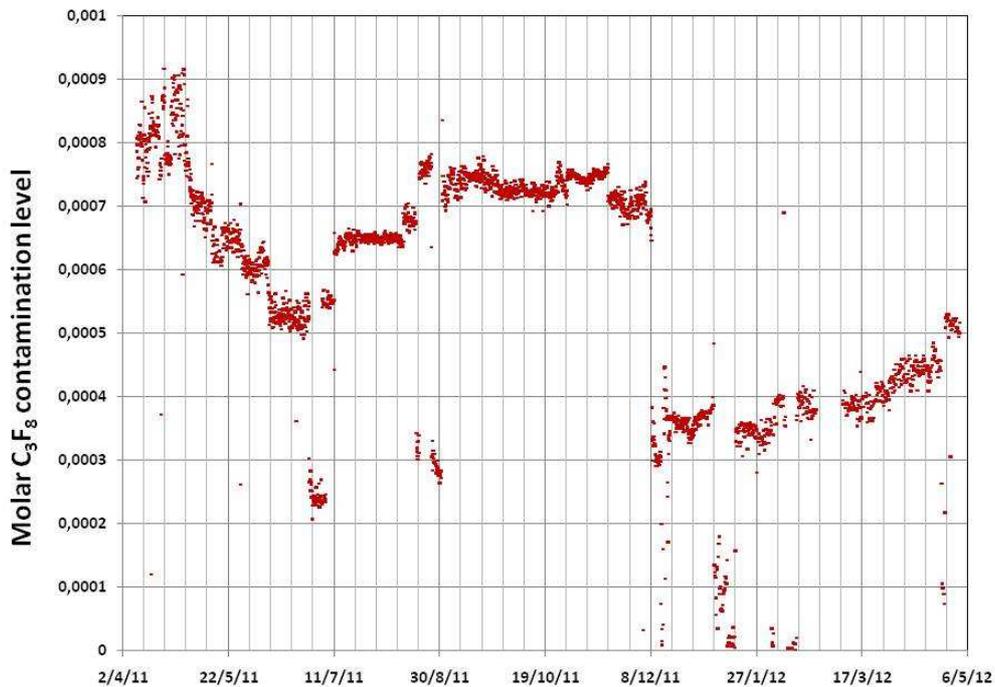

**Figure 12.** Long duration (1 year) log of $C_3F_8$ leak contamination in the $N_2$ environmental gas surrounding the ATLAS pixel detector.

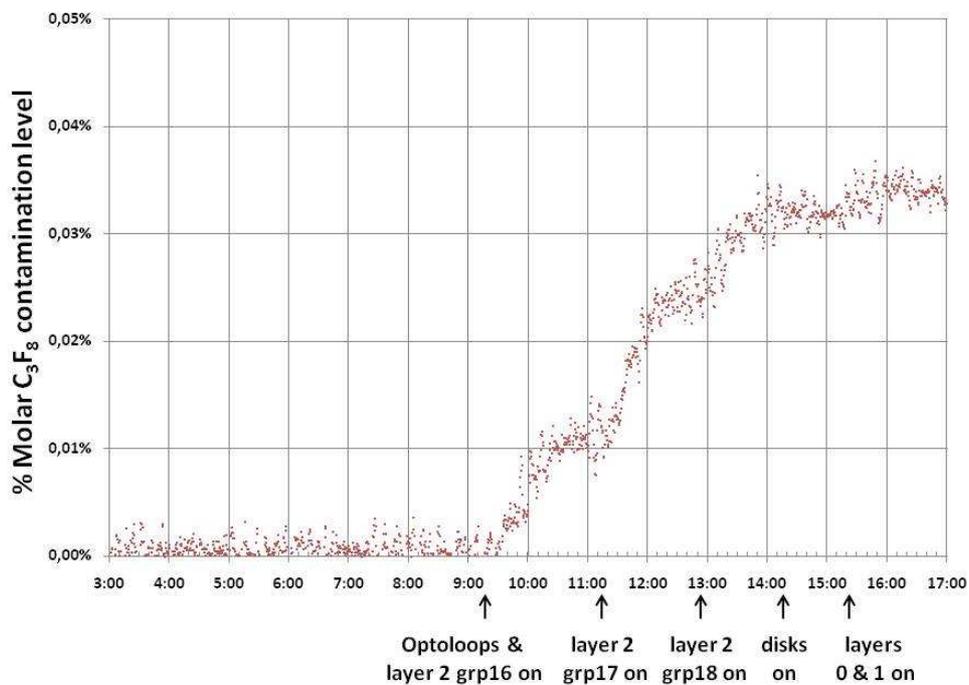

**Figure 13.** The increase of $C_3F_8$ leak contamination during the progressive turn-on of cooling circuits of the ATLAS pixel detector, January 16, 2012.

**3.3 Angled crossing geometry for analysis and in-line flowmetry at high mass flow**

The "angled crossing" geometry ultrasonic flow meter / analyzer is being developed for use at the high fluorocarbon vapour mass flow rate of around 1.2 kgs$^{-1}$ in the principal vapour return tube of the 60kW ATLAS fluorocarbon evaporative cooling system. The present system [1] using compressors to develop the relatively high $C_3F_8$ condensation pressure of 17 bar$_{abs}$ will soon be replaced by a thermosiphon system (section 6) in which fluorocarbon vapour will be condensed at the surface: the liquid delivery pressure being generated hydrostatically in the descent of the 92 m ATLAS pit. The installation of a high flow capacity ultrasonic flow meter/analyzer is foreseen in the superheated vapour return path to the surface, at a location where the vapour will have a probable temperature of 20 ºC and pressure of 500 mbar$_{abs}$. Here a volumetric flow of 307 ls$^{-1}$ is foreseen in pure $C_3F_8$: (density 3.901 kgm$^{-3}$) or 335 ls$^{-1}$ in a blend containing 70 %$C_3F_8$/30 %$C_2F_6$ (density 3.581 kgm$^{-3}$).

In its simplest implementation, an angled ultrasonic flow meter consists of a pair of transducers of diameter *d* separated by a distance *L*, and aligned on a sound path intersecting the main tube, of internal diameter $D_{Main}$, at an angle $\alpha$. The vapour flow rate is calculated from the opposed sound transit times, $t_{down}$, and $t_{up}$, according to the following algorithm:

$$t_{down} = \frac{L}{(c+v\cos\alpha)}, \quad t_{up} = \frac{L}{(c-v\cos\alpha)} \tag{6}$$

where *v* is the linear flow velocity (ms$^{-1}$), *c* the speed of sound in the gas and the distance *L* between transducers is entirely contained within the main flow tube of diameter, $D_{Main}$- implying $L \leq D_{Main}/\sin\alpha$.

The gas volume flow *V* (m$^3$s$^{-1}$) is inferred from the two transit times by:

$$V = \frac{L.\pi.D_{Main}^2(t_{up}-t_{down})}{8\cos\alpha(t_{up}.t_{down})} \tag{7}$$

while the sound velocity *c* can be inferred from the two transit times via:

$$c = \frac{L(t_{up}+t_{down})}{2(t_{up}.t_{down})} \tag{8}$$

If the transducers are attached to the inner wall of the main tube they will impinge on the flow, creating eddies that will affect the measured gas flow velocity. A preferable geometry has the transducers backed-off an additional distance *L'* (counting both sides) to position their inner edges flush with the internal surface of the main tube. This minimal non-impinging transducer spacing can be defined as:

$$L = \frac{D_{Main}}{\sin\alpha} + L', \text{ where } L' = \frac{d}{\tan\alpha} \tag{9}$$

Alternatively the transducers may be withdrawn a longer distance *L'* with respect to the internal surface of the main tube. The latter configuration with a pair of transverse tube stubs allows the changing of a transducer without interruption of the main gas flow: the transducers can be backed off outboard of quarter-turn ball valves, which when open allow the sound path to traverse the main tube.

In configurations where the transducer spacing includes an element *L'* containing quasi-static vapour in communication with the flow in the main tube the formalism for calculating the flow velocity is modified from that of eq. (6) as follows:

$$t_{down} = \frac{L'}{c} + \frac{D_{Main}}{\sin\alpha(c+v\cos\alpha)}, \quad t_{up} = \frac{L'}{c} + \frac{D_{Main}}{\sin\alpha(c-v\cos\alpha)} \tag{10}$$

The sound velocity, *c,* is the physical root derived from the relations of eq. (10) in terms of measurables $L, D_{Main}, \alpha, t_{up}$ and $t_{down}$ :

$$c = \frac{(t_{up}+t_{down})\left(2L'+\frac{D_{Main}}{\sin\alpha}\right) \pm \sqrt{(t_{up}+t_{down})^2\left(2L'+\frac{D_{Main}}{\sin\alpha}\right)^2 - \left(16L't_{up}t_{down}\left(L'+\frac{D_{Main}}{\sin\alpha}\right)\right)}}{4t_{up}t_{down}} \tag{11}$$

allowing the gas flow velocity, *v*, to be calculated as

$$v = \frac{c\left(ct_{up} - \frac{D_{Main}}{\sin\alpha} - L'\right)}{\cos\alpha\,(ct_{up} - L')} \tag{12a}$$

or as

$$v = \frac{c\left(\frac{D_{Main}}{\sin\alpha} + L' - ct_{down}\right)}{\cos\alpha\,(ct_{down} - L')} \tag{12b}$$

CFD simulations were made using the *OpenFOAM®* package [9], [10], [12] to investigate the interaction between the vapour flow in the main tube and the placement of the ultrasonic transducers. The interaction of the flow with the static vapour in the two tube stubs of total length L' was also of interest since the transmitted sound pulse might expected to be disturbed by the presence of vortices in the tube stubs.

Each of the following transducer spacing configurations illustrated in figure 14a-c was evaluated:

- $L = 3D_{main}/\sin\alpha$ : transducers far from the main tube to minimize turbulence effects and allow for valve isolation - *figure 14a*;
- $L = D_{main}/\sin\alpha + d/\tan\alpha$: transducer edges flush with the inner wall of the main tube - *figure 14b;*
- $L = D_{main}/\sin\alpha$: half of each transducer face (diameter, $d$, = 44 mm) impinging on the flow in the main tube - *figure 14c*.

In all cases, the flow conditions were chosen to correspond to those in the main vapour return of the future thermosiphon recirculator, with a mass flow of 1.2 kgs$^{-1}$ of $C_3F_8$ at 20 °C and a pressure of 500 mbar$_{abs}$, (density 3.901 kgm$^{-3}$).

In each transducer configuration simulations were made with sound tube crossing angles of 15°, 30° and 45°. In each case two alternatives were considered for the main tube internal diameter of 133.7 and 211.6 mm (corresponding to the two most probable choices, using standard tube sizes), with corresponding $C_3F_8$ average linear flow velocities of 21.92 and 8.75 ms$^{-1}$. In addition three simulations were made, corresponding to the three crossing angled for half flow in a main tube of internal diameter 133.7 mm.

The number of cells used in the simulation mesh varied between 3 and 8 million, depending on the geometry. The SST *k-ω* [10] turbulence model was adopted in the CFD simulations. Since the fluid flow velocity is lower than Mach 0.3 the resulting variation of density within the system is negligible, allowing the flow in the tube to be modelled as incompressible.

Figure 15 shows negligible influence of the sound tube stubs on the main stream for the case of maximum *(3D$_{Main}$/sinα)* transducer spacing, corresponding to figure 14a. The stream lines remain parallel to each other even close to the stub intersections. Due to the drag force of the main stream on the quasi-static vapour in the sound tube stubs, a slow counterclockwise-rotating vortex occurs in the upper (downstream) stub, while a clockwise-rotating vortex occurs in the lower (upstream) stub. Despite these vortices, the average vapour speed in the two sound tube stubs must be zero, from conservation of mass. It can be seen that a wake follows the downstream tube stub, although this does not influence the flow measurement as it lies outside the sound path. The result is a rather uniform axial velocity distribution across the diameter of the main tube, as shown in figure 16.

The area of the elliptical aperture, $a$, formed by the intersection of the inner diameter, $D_s$, of transversal sound tube stubs and the main tube is given by:

$$a = \frac{D_s^2}{4\sin\alpha} \tag{13}$$

.

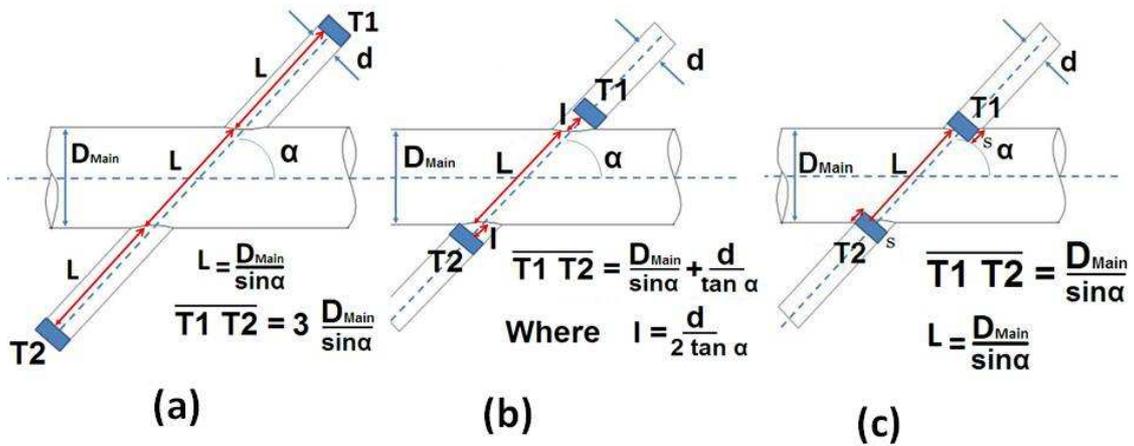

**Figure 14.** Angled flow meter configuration with: (a) 'maximum' transducer spacing $(3*D_{Main}/\sin\alpha)$.
(b) 'mid' transducer spacing $(D_{Main}/\sin\alpha + d/\tan\alpha)$ where $l1 = l2 = d/2\tan\alpha$.
(c) 'minimum' transducer spacing $(D_{Main}/\sin\alpha)$, with transducers impinging on the main flow.

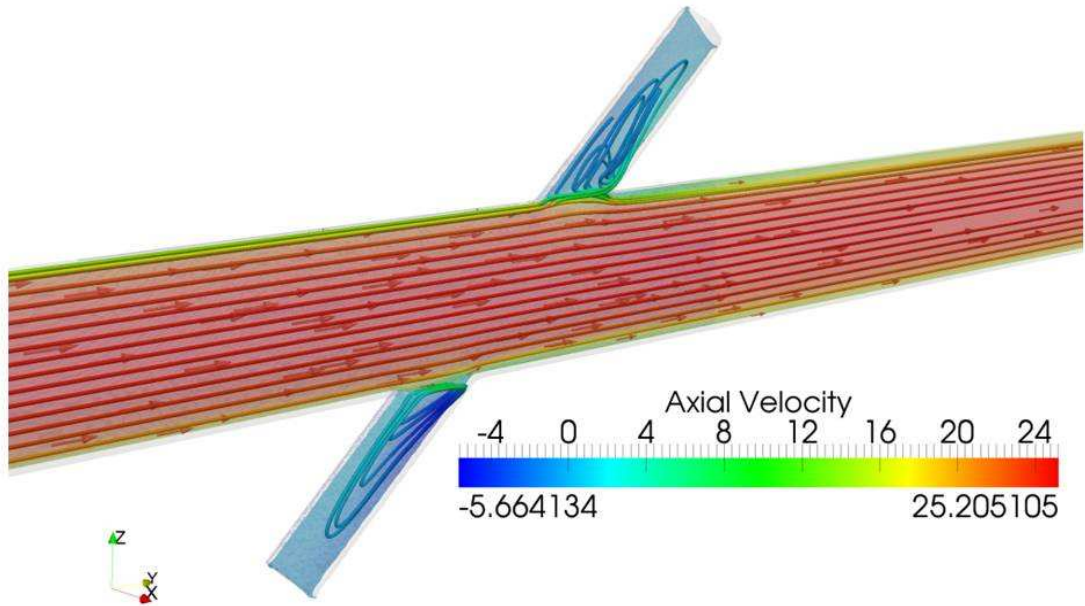

**Figure 15.** Contours of the axial component of gas flow velocity (ms$^{-1}$) on the symmetry plane with stream lines: transducer spacing $(3D_{Main}/\sin\alpha)$, 45° intersection and $D_{main}$ = 133.7 mm.

The effect of elliptical aperture variation can be seen in figure 16. At the shallower crossing angles of 15º and 30º the larger aperture is manifested as a slight increase in the effective diameter of the main tube at the aperture of the downstream sound tube stub. At 45º crossing angle, however, the effective diameter corresponds closely to that of the main tube. Simulations were not therefore carried out at crossing angles exceeding 45º since the $(t_{up}-t_{down})$ transit time difference is progressively reduced as $\alpha$ tends toward 90º, resulting in reduced accuracy in flow measurement. Following the conclusion of prototype tests, discussed later, a crossing angle of 45º was adopted for the final instrument, this being mechanically easier to fabricate in a dissimilar diameter tube cross than the more oblique angles.

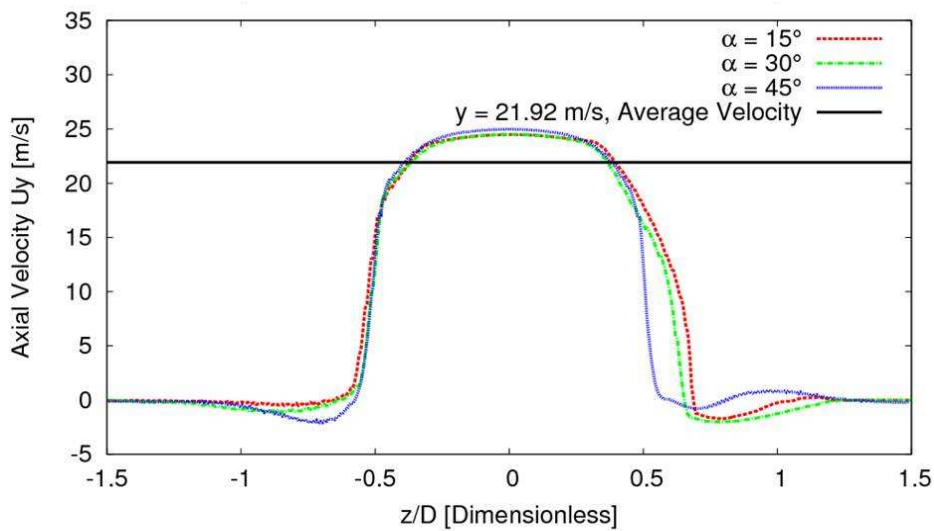

**Figure 16**. Axial flow velocity in (and beyond) the main tube *vs.* position along the sound path between the transducers in the configuration with *($3D_{Main}/sin\alpha$)* spacing. The position along the sound path is a dimensionless projection perpendicular to the axis of the main tube and normalised to $D_{main}$. The average velocity in the 133.7 mm diameter main tube is 21.92 ms$^{-1}$.

The flow profiles in the main tube for the transducer configurations of figures 14a and 14b are similar, (as shown in figures 15 and 17) and exhibit no major influence of the transducers over the main stream flow. On the other hand, in the configuration of figure 14c half the surface area of each transducer is immersed in the main stream. These obstructions create vortices in the sound path near the transducers (figure 18), reducing the accuracy of sound transit time difference measurements. For this reason we have not pursued an internally-mounted transducer configuration through to prototyping.

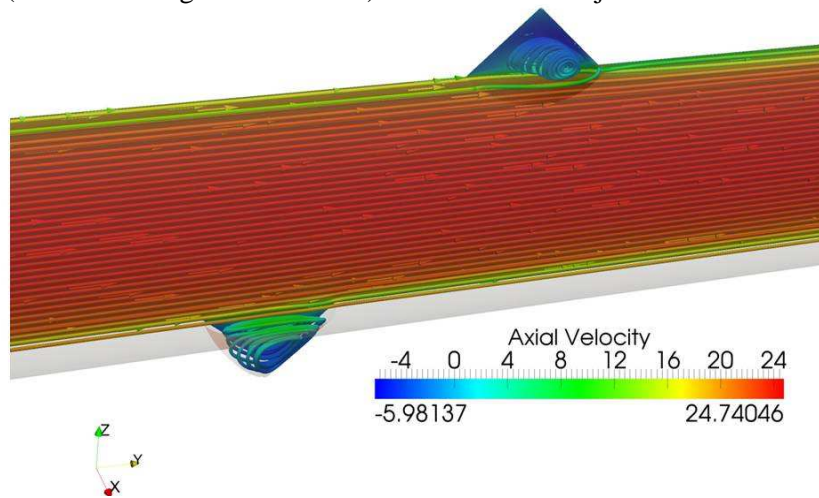

**Figure 17.** Contours of the axial component of velocity on the symmetry plane (ms$^{-1}$), with stream lines: transducer spacing *($D_{Main}/sin\alpha + d/tan\alpha$)*, 45° crossing and $D_{Main}$ = 133.7 mm.

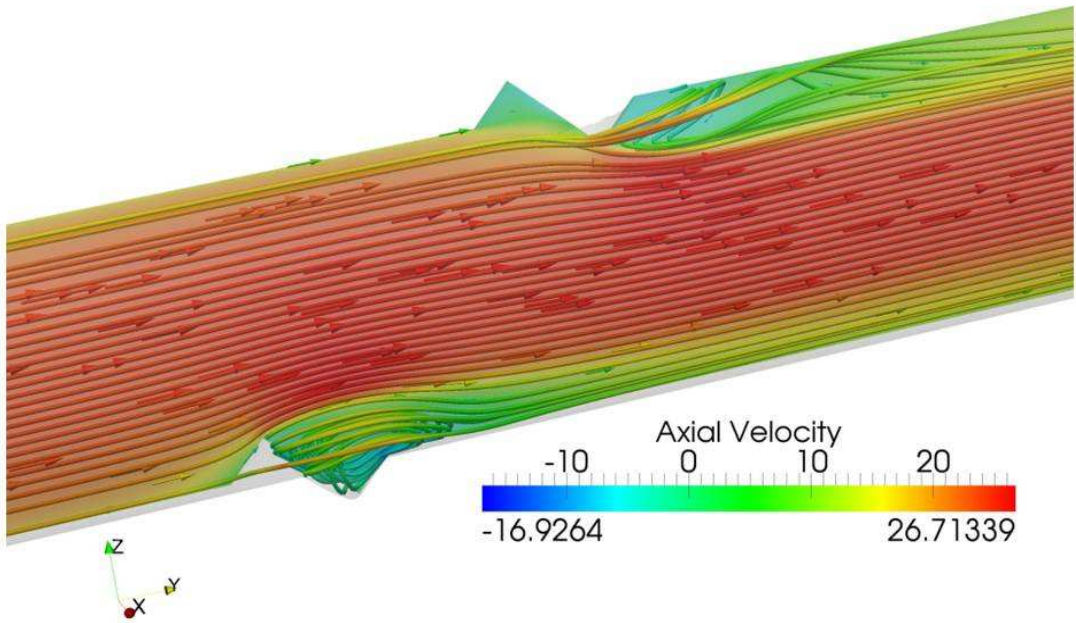

**Figure 18.** Contours of the axial component of velocity on the symmetry plane (ms$^{-1}$) with stream lines: transducer spacing *($D_{Main}/\sin\alpha$)* with 45° crossing and $D_{Main}$ = 133.7 mm.
Roughly half the surface area of each transducer impinges into the flowing vapour.

Table 3 compares the CFD-reconstructed average axial flow velocity (column 3) with the "start condition" average flow in the main tube (column 2) for the 21 different geometrical and flow combinations that were simulated. Geometric correction factors for the different fractional lengths of the sound path in the flowing gas are also evaluated (column 4). The results for the two different main tube diameters show no major influence of this parameter on the calculated flow except in the case (figure 14c) where the transducers impinge on the flow in the main tube. While the choice with $D_{Main}$ = 211.6 mm would allow for a lower pressure drop, the choice with $D_{Main}$ = 133.7 mm is preferable for measurement accuracy, since the velocity component parallel to the sound wave is greater. An inner diameter of 133.7 mm has been subsequently and independently chosen for the local tubing where the ultrasonic flowmeter/analyzer will be installed.

The CFD-calculated average velocity component, $V_c$, in the direction of flow in the main tube (table 3 column 3) is averaged in a cylindrical volume defined by the diameter and spacing of the two transducers, using axial velocity values at points along the whole sound path, including, where applicable, the regions beyond the confines of the main flow tube. The signed percentage deficit of the calculated flow, $V_c$, from the average "starting value" axial flow, $v$, in the main tube, is given by *(($V_c - v$)/$v$)*, and is shown in column 4 of table 3, where it is also compared with the simple expected geometrical deficit, ***GD***. ***GD*** can be expressed as a percentage based on the ratio of the sound path crossing the flowing vapour, ($D_{Main}/\sin\alpha$), to the total length, *L*, including the path length, L', in quasi-static vapour:

$$GD = \frac{100\left(\frac{D_{Main}}{\sin\alpha} - L\right)}{L} \tag{14}$$

Taking an example for the configuration of figure 14a with 15° crossing angle in a 137.1 mm diameter main tube, a value of -64.4 % for $((V_c – v)/v)$ is calculated from CFD. In this geometry 2/3 of the sound path is characterized by quasi-static flow. The expected geometric deficit from sound path arguments according to eq. (14) would be -66.6 %, meaning that the flow velocity measured in such an instrument would need to be multiplied by a geometrical correction factor, *GCF*, to find the axial flow velocity in the main tube, where (maintaining the sign convention of the expected geometric deficit) *GCF* is expressed in terms of *GD* as:

$$GCF = \left(\frac{GD}{100} + 1\right)^{-1} \quad (15)$$

In the configuration of figure 14a, *GCF* = 3, while in the configuration of figure 14b *GCF* depends explicitly on the crossing angle and main tube diameter, as shown in column 5 of table 3. In the configuration of figure 14c, where the entire sound path is in the flowing vapour, *GCF* is unity.

**Table 3.** Comparison of CFD-calculated average axial flow velocities in the inter-transducer sound volume with "starting value" average axial flow in the main tube. Deficits and expected geometric correction factors are shown. Calculations made at half nominal flow are shown in italics.

| $D_{Main}$ (mm), α (°) transducer config. (from figure) | Average axial flow velocity $v$, in main tube (ms$^{-1}$) | CFD volume-averaged flow velocity, $V_c$, inter-transducers (ms$^{-1}$) | CFD calculated % axial flow velocity deficit (expected % geometric deficit *GD*) | Geometric (sound path) correction factor to $V_c$ | Interface aperture (cm$^2$) (5cm ID sound tube) |
|---|---|---|---|---|---|
| 133.7, 15°, (14a) | 21.92 | 7.81 | -64.4 % (-66.6 %) | 3 | 75.9 |
| 133.7, 30°, (14a) | 21.92 | 7.50 | -65.8 % (-66.6 %) | 3 | 39.3 |
| 133.7, 45°, (14a) | 21.92 | 8.05 | -63.3 % (-66.6 %) | 3 | 27.8 |
| 133.7, 15°, (14b) | 21.92 | 17.71 | -19.2 % (-24.1 %) | `1.32 | 75.9 |
| 133.7, 30°, (14b) | 21.92 | 17.68 | -19.3 % (-22.2 %) | 1.29 | 39.3 |
| 133.7, 45°, (14b) | 21.92 | 17.43 | -20.5 % (-18.9 %) | 1.23 | 27.8 |
| 133.7, 15°, (14c) | 21.92 | 19.29 | -12.0 % (0 %) | 1 | 75.9 |
| 133.7, 30°, (14c) | 21.92 | 18.17 | -17.1 % (0 %) | 1 | 39.3 |
| 133.7, 45°, (14c) | 21.92 | 19.00 | -13.3 % (0 %) | 1 | 27.8 |
| *133.7, 30°, (14a)* | *10.96* | *3.76* | *-65.7 % (-66.6 %)* | *3* | *39.3* |
| *133.7, 30°, (14b)* | *10.96* | *8.80* | *-19.7 % (-24.1 %)* | *1.29* | *39.3* |
| *133.7, 30°, (14c)* | *10.96* | *9.20* | *-16.0 % (0 %)* | *1* | *39.3* |
| 211.6, 15°, (14a) | 8.75 | 2.95 | -66.3 % (-66.6 %) | 3 | 75.9 |
| 211.6, 30°, (14a) | 8.75 | 2.84 | -67.5 % (-66.6 %) | 3 | 39.3 |
| 211.6, 45°, (14a) | 8.75 | 2.85 | -67.4 % (-66.6 %) | 3 | 27.8 |
| 211.6, 15°, (14b) | 8.75 | 7.72 | -11.8 % (-16.7 %) | 1.20 | 75.9 |
| 211.6, 30°, (14b) | 8.75 | 7.64 | -12.7 % (-15.3 %) | 1.18 | 39.3 |
| 211.6, 45°, (14b) | 8.75 | 7.71 | -11.8 % (-12.8 %) | 1.15 | 27.8 |
| 211.6, 15°, (14c) | 8.75 | 8.20 | -6.3 % (0 %) | 1 | 75.9 |
| 211.6, 30°, (14c) | 8.75 | 8.03 | -8.3 % (0 %) | 1 | 39.3 |
| 211.6, 45°, (14c) | 8.75 | 8.16 | -6.8 % (0 %) | 1 | 27.8 |

The difference between the CFD-calculated average axial velocity deficit and the expected geometrical deficit (table 3 column 4) is a quality estimator for the particular geometry of ultrasonic flow meter, taking into account the effects of turbulence and vortices in the sound path which are only accessible via the CFD analysis. It can be seen from table 3 that the differences are minimised in the geometry of figure 14a where the transducers are backed off a significant distance from the main tube. In this geometry the average deficit was (65.77 ± 1.52)

%. The standard deviation of the deficits calculated from the 7 independent CFD analyses is thus a reasonable indicator of the precision of CFD predictions.

It can also be seen (table 3 column 6) that the elliptical interface area of the intersecting sound tubes (or the transducer and its support in the geometry of fig 14c) has little influence on the CFD reconstructed average axial velocity for crossing angles between 15 and 45º.

While most of the data in table 3 were calculated at the nominal fluorocarbon mass flow rate of 1.2 kgs$^{-1}$ expected in thermosiphon operation at the full cooling capacity of 60 kW, several simulations were carried out at half-flow in a tube of 133.7 mm internal diameter and a sound path angle of 30° for each of the three transducer configurations. These studies were aimed to check for possible variations of the velocity distribution with the flow rate in the main tube, which might make it difficult to calibrate the instrument. These results are shown in italics in table 3 while the axial flow velocity profile across the main tube is compared with that at full flow in figure 19. The similarity of the flow profiles and the expected deficits - calculated according to eqs. (14) and (15) – for 30° crossing angle at full and half flow leads us to believe that the variations in turbulence within the sound path are insignificant over the flow range expected, and that the calibration of the instrument against, for example, a volumetric flow meter should be linear.

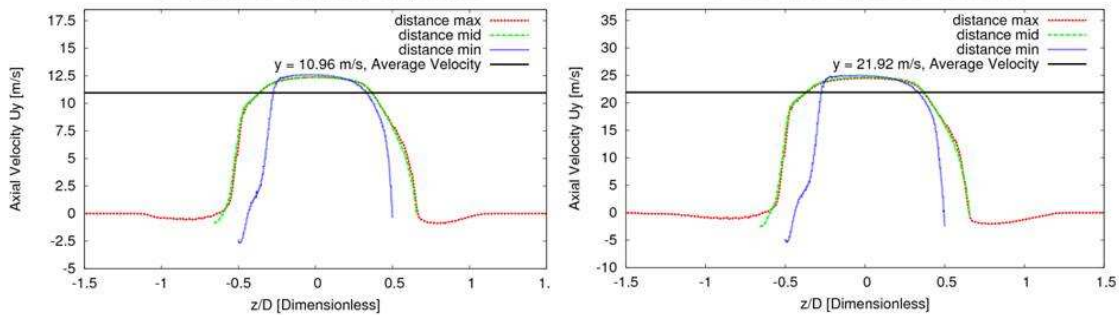

**Figure 19**. Axial flow velocity in (and beyond) the 133.7 mm diameter main tube *vs*. position along the sound path between the transducers in the three configurations for 30° crossing angle: **right:** at full flow of 21.92 ms$^{-1}$; **left**: at half flow of 10.96 ms$^{-1}$. The position along the sound path is a dimensionless projection perpendicular to the axis of the main tube and normalised to $D_{main}$.

In conclusion, CFD simulations have shown that the angled flow meter with transducers withdrawn from the main flow stream can provide a reliable measurement of the average flow velocity, since:
- the sound path covers the whole velocity profile of the main stream;
- the device does not significantly influence the flow;
- vapour in the intersecting sound tube stubs is almost static and the vortices in them are slow;
- calibration is straightforward as the velocity distribution is uniform also for different flow rates;
- the slight turbulent effects do not appear to degrade the sound path over the expected flow range;
- a crossing angle of 45° is preferred.

We have constructed and tested a prototype ultrasonic flowmeter (figures 20, 21) with 45° crossing angle using PVC tubing of similar diameter to that foreseen in the ATLAS thermosiphon installation. The flowmeter was based on a sound tube of 45 mm inner diameter crossing a main tube of 103 mm inner diameter, integrated into an air 'cannon' of total length 7m. The total sound path length was 695 mm, with 550 mm being in static gas in the tube stubs.

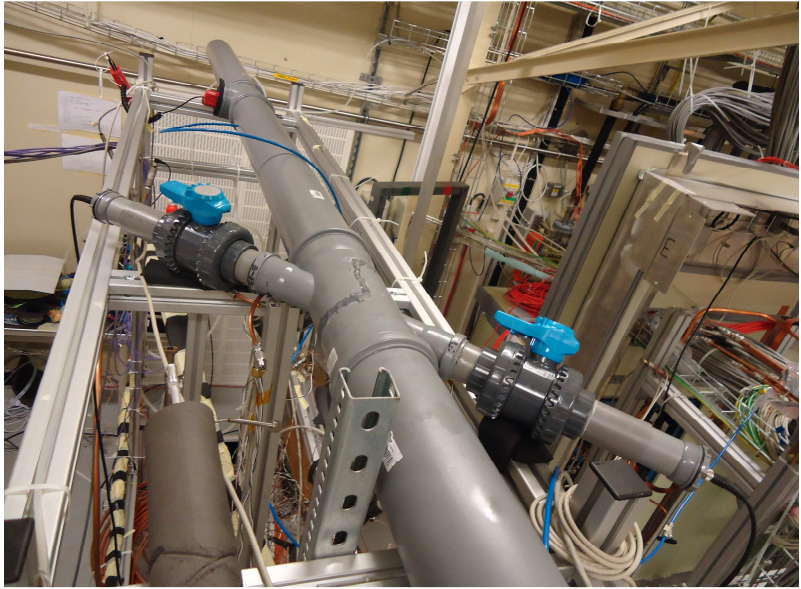

**Figure 20.** Prototype angled ultrasonic flowmeter with 45° crossing angle implemented in PVC tubing of 103mm internal diameter.

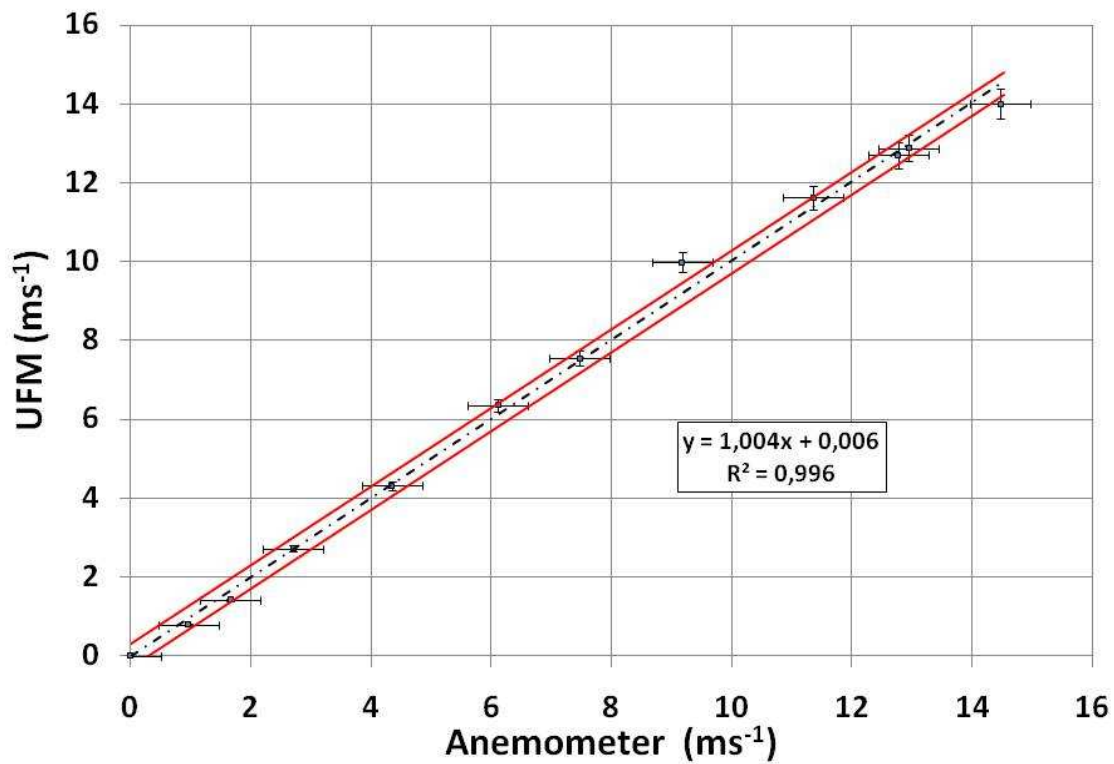

**Figure 21.** Calibration of the PVC prototype 45° angled ultrasonic flowmeter *vs.* anemometer (Amprobe TMA10A). The *rms* deviation of the measured points from the fit represents ± 1.9% of the full flow of 15ms$^{-1}$.

The extra length of the sound tube stubs included the lengths of quarter turn ball valves to allow testing of a geometry similar to the final instrument allowing transducer changeover without stopping the flowing fluorocarbon vapour.

Air was injected from a standard 8 bar building delivery network via a venturi, which could aspirate additional room air for a total maximum flow around 16 ms$^{-1}$. The ultrasonic flowmeter was calibrated against an anemometer (Amprobe model TMA10A: 25 ms$^{-1}$ full scale, intrinsic accuracy ± 2 % FS) installed downstream, as shown in figure 20. The gas flow velocity in the ultrasonic flowmeter was calculated according to the formalism of eqs. 11 and 12a,b. The horizontal error bars in figure 21 represent the quoted 2 % of full scale error of the anemometer. The vertical error bars reflect the combination of the uncertainty in sound velocity measurement (± 0.05 ms$^{-1}$), crossing angle in the PVC tube (± 1 °), PVC main tube diameter (± 1 mm), timing precision (± 100ns) and transducer spacing (± 0.1 mm following length calibration in a quasi-ideal static gas, according to the procedure discussed in section 5). The *rms* accuracy of the ultrasonic flowmeter relative to the fit (shown as red bands in figure 21) is equivalent to ± 1.9 % of the full scale flow of 15 ms$^{-1}$ achievable with the available compressed air supply.

Although calibration with expensive fluorocarbons and their mixtures was not possible in an open system, the encouraging test results have permitted the construction to start on a full scale stainless steel angled ultrasonic flowmeter with 45º crossing angle and a main tube of 137 mm inner diameter.

## 4. Electronics

The custom electronics - shown in figure 22a,b - is based on a Microchip ® dsPIC 16 bit microcontroller. This generates the 50 kHz sound burst signals emitted by the transducers and includes a 40 MHz transit clock that is stopped when an amplified sound signal from a receiving transducer crosses a user-definable comparator threshold. Although the electronics is normally configured for bidirectional 'straight through' transit time measurements with a pair of transducers, calibration operations in echo mode are also possible as in the case of the angled flowmeter geometry, as discussed in section 5.

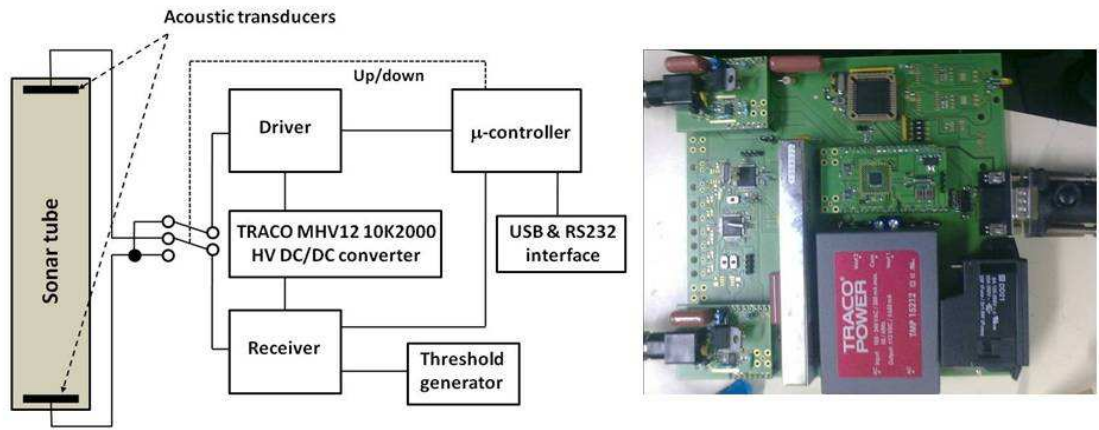

**Figure 22** Left: Block diagram of the bidirectional time-of-flight measuring electronics
Right: Ultrasonic time-of-flight measuring electronics: implementation.

The HV bias for the vibrating foils of the capacitative transducers, settable in the HV range 180-360V, is generated by a DC-DC converter. When transmitting, a transducer is excited with a train of (1-8) HV square wave pulses, built using the 50 kHz LV pulses from the microcontroller and the DC-DC converter output. When receiving, a transducer is biased with a flat HVDC bias and its signal passed to an AD620N amplifier followed by a comparator.

Sound transit times are computed from the number of 40MHz pulses counted between the rising edge of the first transmitted 50 KHz sound pulse, and the time the first received, amplified sound pulse crosses the comparator threshold. Transit times, computed alternately in the two transmission directions are continuously entered into an internal FIFO memory. When a measuring cycle is requested by the supervisory computer a time-stamped running average from the 300 most recent transit times in each direction in the FIFO memory is output, together with the average temperature and pressure, at a rate of up to 20 averaged samples per second.

In addition to the I/O connectivity for communications, the ultrasonic transducers, pressure and temperature sensors, two (4-20 mA) analog outputs provide feedback for adjustment of the $C_3F_8/C_2F_6$ mixing ratio by the external gas mixture control system.

The ultrasonic interface board with the DC-DC conversion and energy storage capacitors for the transducer HV excitation is shown on the left of Figure 22b.

Data is taken over a serial bus running under USB and RS232. Future versions will implement CANbus running the CANOpen communication protocol. The Supervisory, Control and Data Acquisition (SCADA) software - implemented as a PVSS-II project [8]- sends various commands to the local electronics to start, stop or reset the instrument, and retrieve data from the FIFO memory.

The analyzing software which processes the data, makes analyses and generates control signals to the electronics is described in section 6.

## 5. Transducer distance calibration for flowmetry and gas mixture analysis

For high precision mixture and flow analysis the uncertainty in the sound flight distance should be minimized. Following transducer installation it is necessary to perform a transducer foil inter-distance calibration. The most convenient method is to calculate this distance using measured sound transit times with the tube filled with a pure gas (or gases) having well-known sound velocity dependence on temperature and pressure. We initially made calibrations of the 'pinched axial' (section 3.1) and low flow analyzer (section 3.2) instruments using xenon, whose sound velocity and mw (175.5 ms$^{-1}$ at 20 ºC, 137 units – figure 1) are closest [13], [14] to those of the fluorocarbon mixtures in the ATLAS application [1], and whose thermo-physical behaviour is that of an ideal gas. Later calibrations demonstrated sufficient precision with nitrogen and argon, which are considerably cheaper and more widely available. The typical precision in transducer inter-distance measured in this way is ± 0.1mm.

As a precursor to the measurements in $C_3F_8/C_2F_6$ mixtures, transducer inter-distance calibration was followed with measurements in pure $C_3F_8$ and $C_2F_6$. The average difference between measured sound velocities and the predictions [15], [16], [17]) was less than 0.04 % in both pure fluids, and was a strong indicator of a correct transducer inter-distance calibration.

A similar transducer inter-distance calibration procedure will be used in the angled high flow ultrasonic analyzer/flow meter. This configuration must allow for the possible replacement of transducers without interruption of the high vapour flow of around 1.2 kgs$^{-1}$. A transducer may be replaced following closure of the quarter-turn ball valve in the transverse sound tube (as illustrated in figure 20). Following transducer re-installation the air is evacuated from the closed stub of tube outboard of the closed ball valve and the calibration gas injected. The electronics is able to operate in echo mode to find the distance between the transducer foil and the equator of the ball in the closed valve. The calibration gas is then evacuated and the valve is opened to bring the new transducer into contact with the process gas.

The total sound path can be measured with the valves open and the flowmeter filled with static calibration gas before the start of operations with fluorocarbon vapour: the distances of the two transducers from the equators of their valves are subtracted from the total path length to find the distance between the outboard equators of the two ball valves. This distance is assumed to be a 'hard parameter', remaining unchanged following any subsequent transducer replacement.

## 6. SCADA and analysis software

The specialized software for the gas analyzer operation - illustrated in figure 23 - is coded as a standalone component in the PVSS II, v3.8 SCADA environment [8], [15]; a standard at CERN.

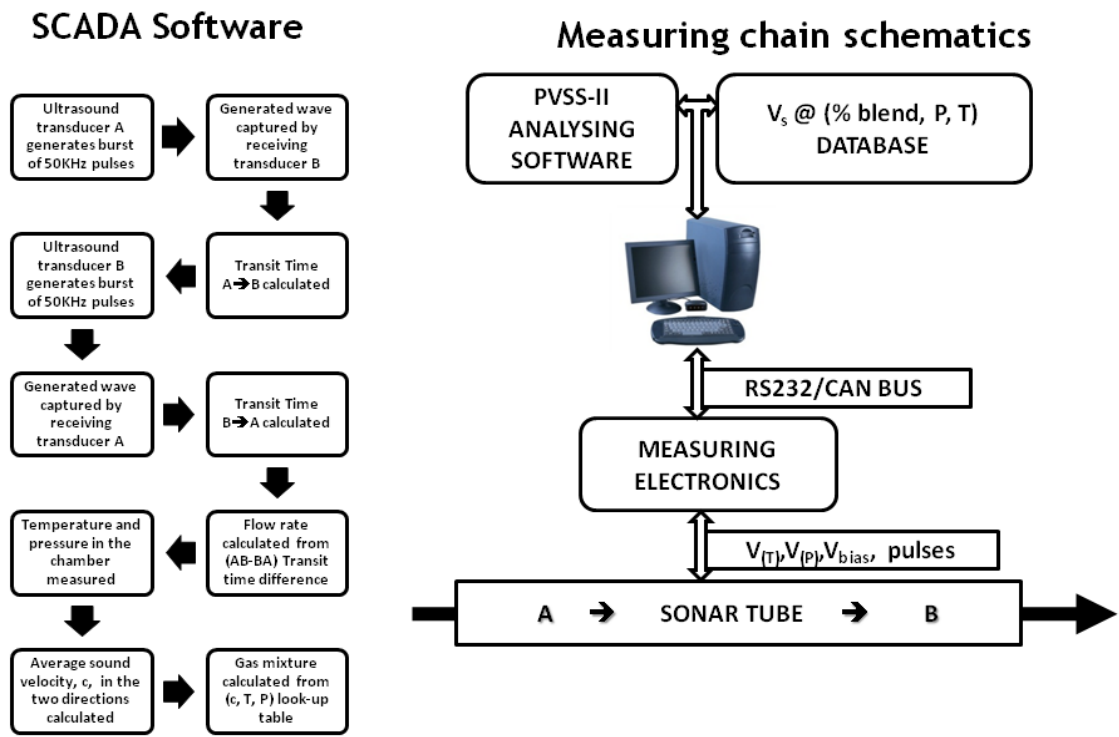

**Figure 23**. The main activities of the PVSS-II SCADA software of the gas analyzer/flow meter.

The main tasks of the software include:

- vapour flow rate determination;
- sound velocity and molar mixture concentration determination;
- communication with the custom electronics, to start and stop the measuring cycle, and to request time-stamped bidirectional sound transit times, temperature and pressure data from the instrument FIFO memory;
- calculation and transmission of the set-points for the analog (4-20mA) output signals for $C_3F_8/C_2F_6$ ratio adjustment in the external cooling plant;
- visualization via a Graphical User Interface (GUI);
- archiving of sound transit times, velocities, flow, mixture composition, temperature and pressure into a local and/or remote data base;

- in the angled flowmeter an echo mode for locating the distance of the transducers from their closed ball valves during calibration with a quasi-ideal gas (section 5).

The present software [15] for $C_3F_8/C_2F_6$ mixture analysis uses a pre-loaded look-up table of NIST-REFPROP generated sound velocity with 0.25 % granularity in $C_3F_8/C_2F_6$ molar mixture and covering the expected range temperature and pressures (16.2➔26.1 °C, 800➔1600 $mbar_{abs}$ with 0.3 °C & 50 mbar granularity). These conditions correspond to the superheated $C_3F_8/C_2F_6$ binary gas mixture environment in the vapour return line where the final instrument will be installed - some tens of metres from the evaporative zone within the silicon tracker.

The algorithm calculates mixture composition by minimizing a quadratic norm, $n_i$, for each ($c_i$, $T_i$, $P_i$) table entry:

$$n_i = k_1(p_{i,table} - p_{running\ average})^2 + k_2(t_{i,table} - t_{running\ average})^2 + k_3(c_{i,table} - c_{running\ average})^2 \quad (16)$$

where $p$, $t$ & $c_{running\ average}$ are real-time outputs of the instrument FIFO memory and $k_{1,2,3}$ are weights [18] that express the dependency of the mixture prediction on the error that comes from the differences between the table and measured $c$, $T$ or $P$ values.

An expanded software version [15] will implement a $C_3F_8/C_2F_6$ database covering a much larger $c$, $T$, $P$ range, from near the critical temperature and pressure down to the expected thermosiphon condenser temperature and pressure of -70 °C and 300 $mbar_{abs}$ (section 7). This development forms part of an ongoing study [7], [13] to feed back improved $C_3F_8/C_2F_6$ thermophysical data into NIST-REFRPOP [5] AND PC-SAFT models. The software will allow "zooming" to smaller sub-tables (O~10,000 $c$, $T$, $P$ data points), corresponding to a narrower process range - as in the present application - in various $C_3F_8/C_2F_6$ mixtures.

The expanded database will also include air/fluorocarbon mixtures over a temperature and pressure range centred on thermosiphon condenser conditions. This database will provide a look up table to determine the level of air ingress into the sub-atmospheric pressure condenser. Look-up table data will be gathered from prior measurements in calibration mixtures and from theoretical data using PC-SAFT equation of state predictions.

## 7. Future ultrasonic vapour analysis developments for the thermosiphons

The present ATLAS silicon tracker $C_3F_8$ evaporative cooling system operates with a condensation pressure of around 17 $bar_{abs}$ [1]. It is impossible to significantly reduce this pressure since the 204 uninsulated $C_3F_8$ liquid supply tubes are grouped parallel and close to electrical cables passing through the ATLAS detector. They are exposed to an ambient temperature of around 40°C, at which the $C_3F_8$ saturated vapour pressure approaches 13 $bar_{abs}$. The requirement that $C_3F_8$ liquid be delivered at a pressure exceeding this value (including a safety factor to prevent premature boiling in the delivery tubing) places severe mechanical constraints on the oil-free compressors used to return $C_3F_8$ vapour to the condensers. The compressors aspirate $C_3F_8$ exhaust vapour from the silicon tracker at a pressure of around 800 $mbar_{abs}$, requiring at a compression ratio in excess of 20. The high compression ratio has proved problematic; compressor failures were frequent in the early days of operation, and while the compressor system is presently (August 2012) operating reliably (with frequent maintenance) the long term-reliability and future mean time between failures (MTBF) continue to cause concern.

An ambitious programme has been undertaken to replace the underground compressors with a thermosiphon circulator [18] in which condensation is carried out on the surface at low pressure and temperature (300 mbar$_{abs}$, -70 ºC), the $\rho gh$ hydrostatic column of around 90 metres of $C_3F_8$ (density around 1.5 kgm$^{-3}$) being used to generate the required liquid delivery pressure. By contrast the hydrostatic column of $C_3F_8$ vapour returning to the surface condenser is only a few tens of mbar. The total flow rate, based on that of the present compressor-based system and a detector cooling power of around 60 kW is around 1.2 kgs$^{-1}$ of fluorocarbon refrigerant. Since it is possible that $C_2F_6/C_3F_8$ blends will be used in the future thermosiphon, studies of the thermal performance of these coolant mixtures have been made in parallel to the work reported in this paper. These thermal evaluations are the subject of a forthcoming publication [19].

In addition to a combined flow meter/analyzer in the vapour return path to the surface condenser, an additional binary mixture analyzer will be connected to the vapour volume (headspace) of the (sub-atmospheric pressure) condenser, as illustrated schematically in figure 24. The sound velocity will be continuously monitored and compared to a velocity/composition look-up table for fluorocarbon/air mixtures. When the sound velocity exceeds that expected in fluorocarbon vapour by more than a pre-defined threshold, the analyzer tube will be isolated from the condenser vapour volume and instead opened to vacuum to eliminate the gas, considered too rich in incondensable air. The frequency of operation of this venting system will depend on the leak rate and volume of the ultrasonic analyzer tube. This system will soon be tested on a small-scale (2 kW, 80 metre depth) thermosiphon installation that has recently been constructed at the ATLAS pit.

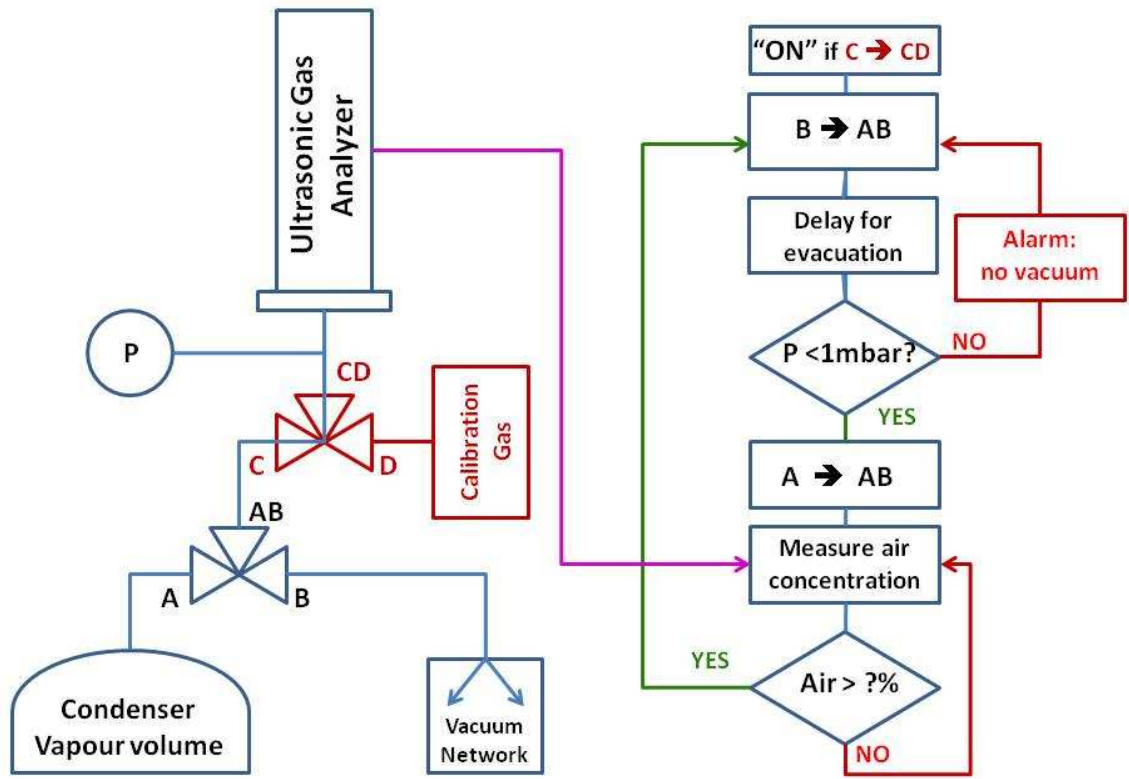

**Figure 24**. Principle of operation of an ultrasonic analyzer used to detect and vent ingressed incondensable gas from a fluorocarbon condenser.

**Conclusion**

We have developed a combined, real-time ultrasonic flow meter and binary gas analyzer with custom electronics and dedicated SCADA software running under PVSS-II, a CERN standard.

One version of the instrument has demonstrated a resolution of $3.10^{-3}$ for $C_3F_8/C_2F_6$ mixtures with ~20 %$C_2F_6$, and a flow precision of 2 % of full scale for fluorocarbon mass flows up to 30 gs$^{-1}$. A second version has been in long term use (over more than 1 year) to monitor $C_3F_8$ leaks into part of the ATLAS silicon tracker nitrogen envelope. Sensitivity to $C_3F_8$ leak concentrations of $< 5.10^{-5}$ has been seen in this instrument.

Several geometries for the instrument are in use or under evaluation. These include a 'pinched axial' sound path geometry for analysis and measurement of moderate flow rates, and an angled sound path geometry for analysis and measurement of high fluorocarbon flow rates of ~1.2 kgs$^{-1}$, corresponding to the full flow in the 60 kW thermosiphon recirculator currently under construction for the ATLAS silicon tracker. Extensive computational fluid dynamics studies were made to determine the preferred geometry, including transducer spacing and placement, together with the sound path crossing angle with respect to the vapour flow direction. A preferred geometry with a main tube inner diameter of 137 mm and a sound crossing angle of 45° has been chosen. Tests in air with a prototype built with PVC tube of similar diameter have demonstrated flow measurement precision of ± 1.9 % of the full scale of 15ms$^{-1}$. The production of a stainless steel variant of the instrument has started and further closed circuits tests with fluorocarbons are planned.

A further variant of the instrument is under development to allow the detection and elimination of incondensable vapour accumulating in the condenser of a fluorocarbon recirculator [20].

The instruments described in this work have many potential applications, including the analysis and flowmetry of hydrocarbons, vapour mixtures for semi-conductor manufacture and anaesthetic gas mixtures.

**Acknowledgments**

The authors wish to thank CERN and their home institutes for support for this project. University of Indiana participation was supported through U.S. Department of Energy contract DE-AC02-98CH10886. University of Oklahoma participation was supported through DOE contract DE-FG02-04ER41305. A. Bitadze acknowledges support for his PhD research from the ATLAS inner detector group and the particle physics group of the University of Glasgow. R. Bates acknowledges support from the UK Science and Technology Facilities Council. M. Doubek, V. Vacek and M. Vitek acknowledge individual support and support for this project through the following grants in the Czech Republic: MSM grants 6840770035 & LA08015 and SGS/FIS grant10-802460 of the Czech Technical University, Prague.